\documentclass[aps,amsmath,amssymb,superscriptaddress,prb,reprint,twocolumn,longbibliography]{revtex4-2}
\usepackage{graphicx}
\usepackage{dcolumn}
\usepackage{braket,bm}

\usepackage{mathptmx}

\begin{document}

\preprint{APS/123-QED}

\title{Symmetry manipulation of nonlinear optical effect for metallic TMDC}
\author{Ren Habara}
\affiliation{Department of Nanotechnology for Sustainable Energy, School
of Science and Technology, Kwansei Gakuin University, Gakuen-Uegahara 1, Sanda 669-1330, Japan}
\author{Katsunori Wakabayashi}
\affiliation{Department of Nanotechnology for Sustainable Energy, School
of Science and Technology, Kwansei Gakuin University, Gakuen-Uegahara 1, Sanda 669-1330, Japan}
\affiliation{National Institute for Materials Science (NIMS), Namiki 1-1, Tsukuba 305-0044, Japan}
\affiliation{Center for Spintronics Research Network (CSRN), Osaka
University, Toyonaka 560-8531, Japan}

\date{\today}

\begin{abstract}
  Nonlinear optical (NLO) effect plays a crucial role to engineer
 optical angular frequency and symmetry of electronic system. 
  Metallic transition-metal dichalcogenide (TMDC) is one of two-dimensional (2D) materials, which has no inversion symmetry for odd-number-layer.
  In particular, odd-number-layered NbSe$_2$ has spin splitting owing to Ising-type spin-orbit coupling.
  In this paper, we numerically calculate the NLO charge and spin
 conductivities of NbSe$_2$ based on an effective tight-binding model
 for several different optical effects, i.e., symmetry manipulation by
 bi-circular light (BCL) and bulk photovoltaic effect (shift and
 injection current).  
  Under irradiation of BCL which can control the symmetry of electronic system, the current can be generated even in even-number-layered NbSe$_2$.
  Also, we find that shift current can be generated for odd-number-layered NbSe$_2$, which is robust against electronic scattering, i.e., topological current.
  The direction of generated shift current can be switched by altering polarization of light.
  Our results will serve to design opt-spintronics devices based on 2D
 materials to manipulate the charge and spin current and their
 directions by controlling the polarization of incident light which
 recasts the symmetry of electronic system. 
  \end{abstract}

\maketitle

\section{Introduction}\label{sec:level1}
Interaction of strong coherent light with matter 
advanced the field of nonlinear optics, which 
induces a charge polarization nonlinearly. 
The emergence of second-order nonlinear optical (NLO) effect has been
extensively studied in condensed matter 
physics and materials science~\cite{Franken1961, Bolem1996, Boyd2008,
Sipe2000}. 
It provides the foundation of laser frequency conversion~\cite{Wen2019, Trolle2014, Kumar2013, Moss1987, Ghahramani1991, Malard2013, Seyler2015, Rashkeev1998, Rashkeev2001, Sharma2004, Leitsmann2005, Shen1989, Shen2020, Ishiyama2014, Stock2020, Winterfeldt2008, Vampa2014} and direct current (DC) 
photocurrent generation~\cite{Morimoto2016, Nakamura2017, Cook2017,
Fregoso2017, Zhang2019, Zhang20192, Xu2021, Akamatsu2021,
Schankler2021, Ideue2021}. 
In the NLO process, the crystal symmetry is significant.
If a crystal is centrosymmetric, no charge polarization occurs.
In noncentrosymmetric crystal, however, 
nonlinear charge polarizaiton is induced which can generate
DC photocurrent without p-n junctions~\cite{Morimoto2016, Nakamura2017, Cook2017,
Fregoso2017, Zhang2019, Zhang20192, Xu2021, Akamatsu2021, Schankler2021,
Ideue2021} and second-harmonic generation (SHG)~\cite{Wen2019, Trolle2014, Kumar2013, Moss1987, Ghahramani1991, Malard2013, Seyler2015, Rashkeev1998, Rashkeev2001, Sharma2004, Leitsmann2005}.

Transition-metal dichalcogenide (TMDC) 
with the chemical formula MX$_2$ (M = Mo, W, Nb, Ta; X = S, Se) 
is layered materials, 
which can be easily exfoliated
into monolayer due to weak van der Waals forces between layers of TMDC~\cite{Novoselov2005, Desai2016, Lin2016, Yu2018, Wang2021}. 
Thus, TMDC forms a new class of atomically thin two-dimensional
(2D) electronic systems. 
For NLO response, the advantage of atomically-thin 2D materials such as graphene and TMDC
is that the phase-matching conditions
between incident light and light-induced electric polarization wave 
 can be ignored because of
their extremely smaller thickness than the incident light wavelength~\cite{Wen2019, Fryett2017, He2021}.
Thus, in few-layered TMDC the NLO effect mainly depends on the crystal
symmetry.


The generation mechanisms for NLO current are dominated by parity
symmetry $P$~\cite{Xu2021}. 
Under $P$-conserved, NLO charge and spin current is absent.
On the other hand, for $P$-broken, bulk photovoltaic effect is induced in the system.
In general, bulk photovoltaic effect has two contributions: (i) shift
current and (ii) injection current~\cite{Xu2021, Ideue2021, Sipe2000, Fei2020, Bieler2006, Fregoso2019}. 
(i) Shift current is the photoinduced spontaneous DC,
which corresponds to the shift of electrons in real space
during the optical excitaiton of electron. 
(ii) Injection current is net current for different velocities between
electron and hole owing to the population imbalance between them induced
by photo-excitation. 
In the noncentrosymmetric system, the linearly polarized (LP) light
irradiation solely induces the charge shift current, however the incident circularly polarized (CP) light solely induces 
the charge injection current~\cite{Xu2021}. 

In further, bi-circular light (BCL) is useful method to optically
engineer the symmetry of electronic system~\cite{Ikeda2022, Dorney2021, Trevisan2022, JimNat2020, Odzak2015, Pisanty2014}. 
BCL is the superposition of left-handed CP (LCP) light with the angular
frequency $n_1\omega$ and right-handed CP (RCP) light with $n_2\omega$
($n_1\neq n_2$), which forms the trajectory of rose curve. 
It is defined as
\begin{equation}
  A_{\rm{BCL}}(t)=A_Le^{in_1\omega t}+A_Re^{in_2\omega t-i\theta}+{\rm c.c.},
  \label{eq:bclform}
\end{equation}
where $A_{L(R)}$ is amplitude of LCP(RCP) light and $\theta$ is a phase difference between LCP and RCP light.
The application of BCL can artificially control the symmetry of electronic system and induce charge polarizations along the directions of leaves of BCL.

In this paper, we theoretically consider the NLO effects on metallic TMDCs.
The metallic TMDCs 
such as NbSe$_2$, NbS$_2$, TaSe$_2$ and TaS$_2$ are metallic
at room temperature, and successively show
the charge density wave (CDW) phase~\cite{Xi2015, Chatterjee2015}
 and  superconducting phase at low temperature~\cite{Wilson2001, Kim2017, He2018, Xi2016, Sohn2018, Anikin2020, Lian2017}. 
In these materials, AB-stacking structure is most stable in nature and
has different crystal symmetries for even and odd number of layers. 
Even-number-layered TMDCs have a space group D$_{3d}$, which respect to
inversion and out-of-plane mirror symmetries. However odd-number-layered
TMDCs have a space group D$_{3h}$, which break inversion symmetry.
In further, owing to the broken inversion symmetry and a strong atomic spin-orbit
coupling (SOC) field such as Nb and Ta atoms, its system possesses Ising-type
SOC~\cite{He2018, Xi2016, Sohn2018, Lu2015, Saito2016, Zhou2016, Bawden2016}, i.e., an effective Zeeman field that locks electron spins to
out-of-plane directions by in-plane momentum.

In NbSe$_2$, the SOC causes large spin splitting in the energy band structures of
odd-number-layered systems (about $157$ meV at K point), which leads to
unconventional topological spin properties. 
In actual, we have shown that monolayer NbSe$_2$ can generate the spin
Hall current under visible light irradiation owing to its finite topological spin Berry curvature~\cite{Habara2021}.
Also, we have studied that the second-order NLO charge and spin Hall current of SHG process can be selectively generated in few-layered NbSe$_2$ according to the crystal symmetry and polarization of incident light~\cite{Habara2022}.

Here, we extend our theoretical analysis to NLO charge and spin conductivities under BCL irradiation and DC photocurrent (shift and
injection current) in few-layered NbSe$_2$. 
We employ the effective tight-binding model (TBM) including the electron hopping among
$d_{z^2}$, $d_{xy}$ and $d_{x^2-y^2}$ orbitals of Nb atom and Ising-type
SOC in order to describe the electronic structures of NbSe$_2$ around
Fermi level. 
In general, the second-order NLO current is generated in
odd-number-layered NbSe$_2$, but absent in even-number-layered
NbSe$_2$. 
However, since BCL can manipulate the symmetry of electronic states, the
NLO current can be generated even in even-number-layered
NbSe$_2$. 
In further, we find that the charge and spin shift current can be
generated in odd-number-layered NbSe$_2$, which is robust to the
electronic scattering, i.e., topological current. 
In addition, the direction of generated shift current can be switched if LP light is altered to CP light.
Our results will serve to design opt-spintronics devices on the basis of
2D materials to manipulate the charge and spin current and their directions by controlling the polarization of incident light which recasts the crystal symmetry.

This paper is organized as follows.
In Sec.~\ref{sec:level2} we discuss the crystal symmetry, 
electronic structures of few-layered NbSe$_2$ based on the effective TBM. 
In Sec.~\ref{sec:level3} we briefly introduce the formula to calculate
the second-order NLO conductivities and discuss their
relation to the crystal symmetry.
In Sec.~\ref{sec:level4} we will show that incident BCL can control the
symmetry of electronic system, which induces the NLO current along the directions of leaves of BCL.
In Sec.~\ref{sec:level5} we show that the shift current has NLO
selection rule depending on the crystal symmetry and polarization of
incident light. It is also shown that charge and spin shift current can
be switched if the polarization altered. 
Sec.~\ref{sec:level6} provides the summary of our results.
In Appendix, we give the symmetry analysis on NLO conductivity, NLO
current induced by multi-leaf BCL and contour plots of integrand of NLO
conductivities. 
In Supplementary Material, we show the imaginary parts of the NLO
conductivities for BCL, rotational angle dependences of BCL on NLO
conductivities, the derivation of shift- and injection-current
conductivities, and the NLO conductivity for MoS$_2$ as a reference of a TMDC semiconductor~\cite{supplementary}.

\section{Model}\label{sec:level2}
\begin{figure*}[t]
  \begin{center}
    \includegraphics[width=1.0\textwidth]{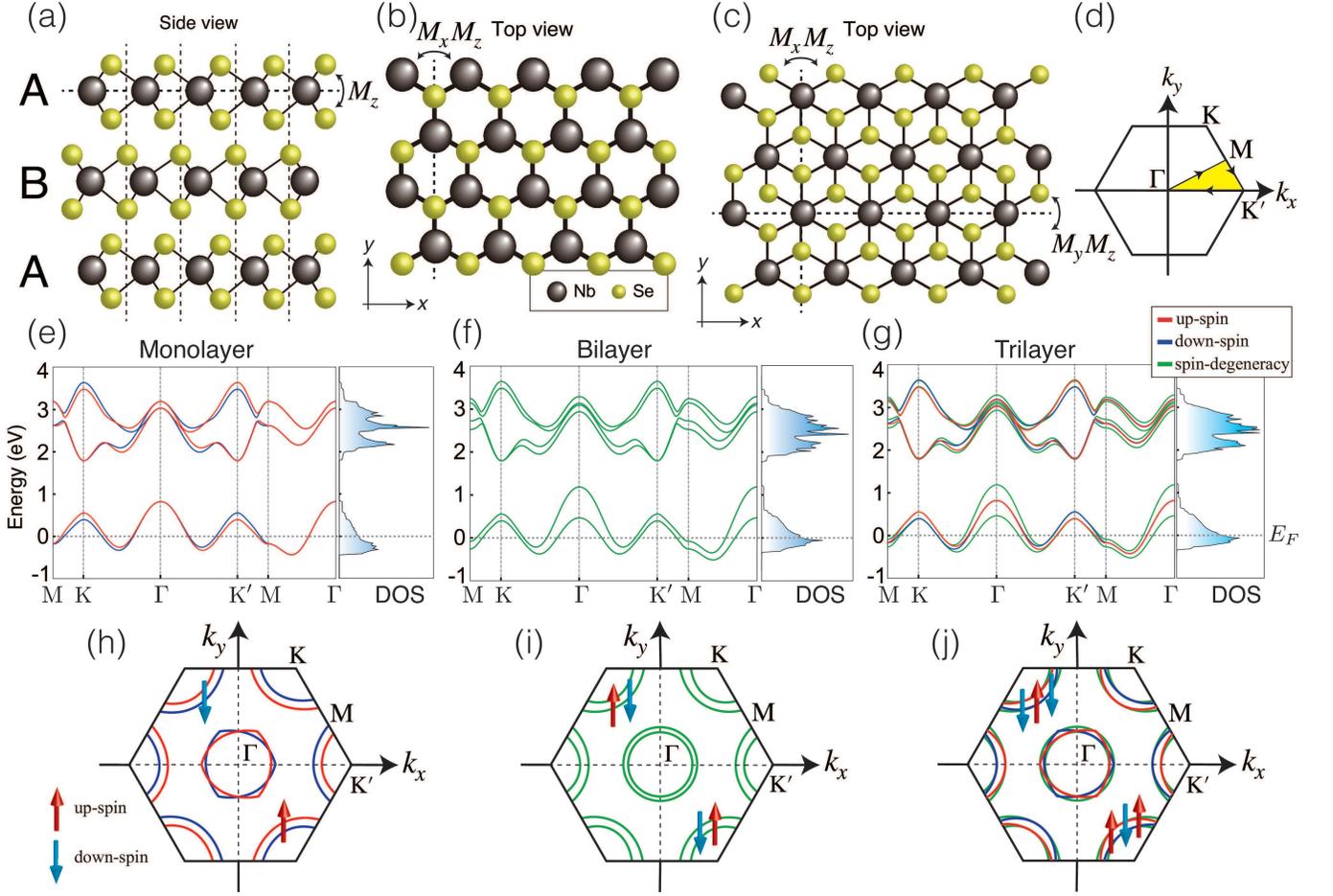}
    \caption{Crystal structures of few-layered NbSe$_2$ with Nb (black) and Se (yellow) atoms.
      (a) Side view of crystal structure of monolayer (A), bilayer (AB) and trilayer (ABA) NbSe$_2$.
      Monolayer NbSe$_2$ has mirror symmetry $M_z$ in perpendicular direction with respect to the plane.
      Top views of crystal structures of (b) monolayer, (c) bilayer and trilayer NbSe$_2$, respectively.
      Odd-number-layered NbSe$_2$ has mirror symmetry $M_xM_z$ and no inversion symmetry, but even-number-layered NbSe$_2$ has inversion symmetry because of mirror symmetries $M_xM_z$ and $M_yM_z$.
      (d) 1st BZ of NbSe$_2$.
      Energy band structures and DOS of (e) monolayer, (f) bilayer and (g) trilayer NbSe$_2$ with SOC parameter $\lambda_{\rm{SOC}}=0.0784$ eV, respectively.
      Fermi level is set to zero.
      Fermi surfaces of (h) monolayer, (i) bilayer and (j) trilayer NbSe$_2$, respectively.
      Red, blue and green lines indicate up-spin, down-spin and spin-degeneracy states, respectively.}
   \label{fig:1}
  \end{center}
\end{figure*}
We will show that the second-order NLO charge and spin conductivities strongly depend on the crystal symmetry and the
polarization of light in metallic TMDC.
In this paper, we employ NbSe$_2$ as an exapmle of metallic TMDCs.
Since NbSe$_2$ can be easily exfoliated because of van der Waals forces between layers, few-layered NbSe$_2$ such as monolayer, AB-stacked bilayer and ABA-stacked trilayer is obtained.
In particular we focus on two cases which are odd-number-layered (e.g. monolayer and ABA-stacked trilayer) and even-number-layered (e.g. AB-stacked bilayer) NbSe$_2$.
Figure~\ref{fig:1} (a) shows the side view of few-layered NbSe$_2$.
The stacking structure is consisted by inserting B monolayer ($180^\circ$ rotation of A monolayer) between A monolayers, which is the most energetically stable stacking sequence in NbSe$_2$.
Each layer (monolayer) has the out-of-plane mirror symmetry $M_z$ in perpendicular direction with respect to the plane of Nb atoms.
Figures~\ref{fig:1} (b) and (c) show the top views of monolayer and AB-stacked bilayer (ABA-stacked trilayer) NbSe$_2$, respectively.
Odd-number-layered NbSe$_2$ has
a space group D$_{3h}$, which has mirror symmetry $M_xM_z$ and no inversion symmetry. 
On the other hand, in even-number-layered NbSe$_2$, the system has a space group D$_{3d}$, which respects to two mirror symmetries $M_xM_z$ and $M_yM_z$ and inversion symmetry $P$.
Figure~\ref{fig:1} (d) shows first Brillouin Zone (BZ) for few-layered NbSe$_2$.

For NbSe$_2$, we employ a multi-orbitals TBM which includes $d_{z^2}$, $d_{xy}$ and $d_{x^2-y^2}$ orbitals of Nb atom, which can well describe the electronic states of NbSe$_2$ around Fermi energy ($E_F=0$ eV)~\cite{He2018, Habara2021, Habara2022, Liu2013, Kim2021}.
The eigenvalue equation for the effective TBM is
\begin{math}
 \hat{H}(\bm{k})|u_{n\bm{k}}\rangle = E_{n\bm{k}}|u_{n\bm{k}}\rangle,
\end{math}
where $n$ is the band index with spin states expressed as $n=1,2,\cdots,6N$ ($N$ is the number of layer), $\bm{k}=(k_x,k_y)$ is the wave-number vector for 2D material and $E_{n\bm{k}}$ is the eigenvalue for $n$-th band.
The eigenvector for $n$-th band is defined as
$|u_{n\bm{k}}\rangle = (c_{n\bm{k},d_{z^2},\uparrow},c_{n\bm{k},d_{xy},\uparrow},c_{n\bm{k},d_{x^2-y^2},\uparrow},c_{n\bm{k},d_{z^2},\downarrow},c_{n\bm{k},d_{xy},\downarrow},c_{n\bm{k},d_{x^2-y^2},\downarrow})^T$,
where $(\cdots)^T$ indicates the transpose of vector and
$c_{n\bm{k}\tau s}$ means the amplitude at atomic orbital $\tau$ with
spin $s$ for the $n$-th energy band at $\bm{k}$.
The Hamiltonian of monolayer NbSe$_2$ with Ising-type SOC is 
\begin{equation}
 \hat{H}_{\rm{mono}}(\bm{k})=\hat{\sigma}_0\otimes \hat{H}_{\rm{TNN}}(\bm{k})+\hat{\sigma}_z\otimes\frac{1}{2}\lambda_{\rm{SOC}} \hat{L}_z
\end{equation}
with
\begin{equation}
 \hat{H}_{\rm{TNN}}(\bm{k})=\begin{pmatrix}V_{0}&V_{1}&V_{2}\\ V_{1}^{*}&V_{11}&V_{12}\\ V_{2}^{*}&V_{12}^{*}&V_{22}\\ \end{pmatrix},\ \hat{L}_z=\begin{pmatrix}0&0&0\\ 0&0&-2i\\ 0&2i&0\\ \end{pmatrix},
\end{equation}
 where $\hat{\sigma}_0$ and $\hat{\sigma}_z$ are Pauli matrices and $\lambda_{\rm{SOC}}$ is the Ising-type SOC parameter ($\lambda_{\rm{SOC}}=0.0784$ eV for monolayer NbSe$_2$).
$\hat{H}_{\rm{TNN}}(\bm{k})$ includes the electron hoppings only among three
 $d$-orbitals of Nb atoms such as $d_{z^2}$, $d_{xy}$ and $d_{x^2-y^2}$,
 which are assumed up to third-nearest neighbor sites. 
Similarly, using the Hamiltonian of monolayer NbSe$_2$ $\hat{H}_{\rm{mono}}(\bm{k})$, the Hamiltonians of bilayer and trilayer NbSe$_2$ can be written as
\begin{equation}
 \hat{H}_{\rm{bi}}(\bm{k})=
  \begin{pmatrix}
   \hat{H}_{\rm{mono}}(-\bm{k})&\hat{H}_{\rm{int}}(\bm{k})\\
   \hat{H}^{\dag}_{\rm{int}}(\bm{k})&\hat{H}_{\rm{mono}}(\bm{k})\\
  \end{pmatrix}
\end{equation}
and
\begin{equation}
 \hat{H}_{\rm{tri}}(\bm{k})=
  \begin{pmatrix}
   \hat{H}_{\rm{mono}}(\bm{k})&\hat{H}_{\rm{int}}(\bm{k})&0\\
   \hat{H}^{\dag}_{\rm{int}}(\bm{k})&\hat{H}_{\rm{mono}}(-\bm{k})&\hat{H}_{\rm{int}}(\bm{k})\\
   0&\hat{H}^{\dag}_{\rm{int}}(\bm{k})&\hat{H}_{\rm{mono}}(\bm{k})\\
  \end{pmatrix},
\end{equation}
respectively~\cite{Sohn2018}. 
Here, the interlayer coupling Hamiltonian $\hat{H}_{\rm{int}}(\bm{k})$ between monolayers is
\begin{equation}
 \hat{H}_{\rm{int}}(\bm{k})=
  \begin{pmatrix}
   T_{01}&0&0\\
   0&T_{02}&0\\
   0&0&T_{02}\\
  \end{pmatrix}.
\end{equation}
The details of matrix elements $V_0$, $V_1$, $V_2$, $V_{11}$, $V_{12}$,
$V_{22}$, $T_{01}$ and $T_{02}$ can be found in Ref.~\cite{Habara2021, Habara2022}.

Figures~\ref{fig:1} (e), (f) and (g) show the energy band structures and
density of states (DOS) of monolayer, bilayer and trilayer NbSe$_2$,
respectively. 
Here, red, blue and green lines indicate up-spin, down-spin and spin-degeneracy states, respectively.
Fermi energy $E_F$ is set to $0$ eV.
Also, the system has a large energy band gap between the partially
filled valence bands and empty conduction bands (about $1.5$ eV). 
Because of the broken inversion symmetry in odd-number-layered NbSe$_2$,
the spin splitting bands can be seen, which show opposite spin states at
the valence band edges at K and K$^{\prime}$ points in the energy band
structure, i.e., preservation of time-reversal symmetry. 
On the other hand, since even-number-layered NbSe$_2$ has crystal
inversion symmetry, spin-degeneracy states appear in the energy band
structure. 
In even-number-layered NbSe$_2$, the interlayer interaction has larger effects on the valence band at $\Gamma$ point than K and K$^{\prime}$ points,
which leads to the large band splitting of valence band at $\Gamma$
point.
In trilayer NbSe$_2$, the energy band consists of spin-splitting and
spin-degeneracy states.
It should be noted that the energy band structure of trilayer NbSe$_2$
can be understood by superposition of monolayer and bilayer NbSe$_2$.

Figures~\ref{fig:1} (h), (i) and (j) show the spin-dependent Fermi
surfaces of monolayer, bilayer and trilayer NbSe$_2$, respectively. 
In odd-number-layered NbSe$_2$, the Fermi surface has up- and
down-spin hole pockets centered at $\Gamma$, K and K$^{\prime}$ points. 
However, in even-number-layered NbSe$_2$, no such spin splitting occurs
owing to the inversion symmetry.
Thus, we show that the spin dependence of Fermi surfaces behaves differently for even- and odd-number-layered NbSe$_2$.

\section{Second-order NLO charge and spin conductivities}\label{sec:level3}
\begin{table*}
  \caption{\label{tab:table1}Second-order NLO charge and spin conductivities of SHG process $\sigma^{\rm{charge}}_{ijk}(2\omega; \omega, \omega)$ and $\sigma^{\rm{spin}}_{ijk}(2\omega; \omega, \omega)$ for odd- and even-number-layered NbSe$_2$~\cite{Habara2022}}.
  \begin{ruledtabular}
    \begin{tabular}{ccc}
      &second-order NLO charge conductivity&second-order NLO spin conductivity \\ \hline
      odd-number-layer (D$_{3h}$)&$\sigma^{\rm{charge}}_{yyy}=-\sigma^{\rm{charge}}_{yxx}=-\sigma^{\rm{charge}}_{xyx}=-\sigma^{\rm{charge}}_{xxy}$&$\sigma^{\rm{spin}}_{xxx}=-\sigma^{\rm{spin}}_{xyy}=-\sigma^{\rm{spin}}_{yxy}=-\sigma^{\rm{spin}}_{yyx}$ \\
      even-number-layer (D$_{3d}$)&zero&zero \\
    \end{tabular}
  \end{ruledtabular}
\end{table*}
Under irradiation of light with strong amplitude, the current density $J_i$ can be expanded as follows, i.e., $J_i=J^{(1)}_i+J^{(2)}_i+J^{(3)}_i+\cdots$.
Here, $J^{(1)}_i$ is the linear optical current.
$J^{(2)}_i$ and $J^{(3)}_i$ are second- and third-order NLO current.
In second-order NLO effect, the generated current density $J^{(2)}_i$ can be given as
\begin{equation}
  J^{(2)}_i=\sum_{jk}\sum_{\omega_1\omega_2}\sigma^{(2)}_{ijk}(\omega_1+\omega_2;\omega_1,\omega_2)E_j(\omega_1)E_k(\omega_2),
  \label{eq:j2}
\end{equation}
where $E_j(\omega_1)$ and $E_k(\omega_2)$ are electric fields of incident light~\cite{Boyd2008}.
Note that $(i, j, k)$ is $x$- or $y$-direction.
In particular, $i$ means the generation direction of NLO current.
$j$ and $k$ are the polarization of $E_j(\omega_1)$ and $E_k(\omega_2)$.
$\sigma^{(2)}_{ijk}(\omega_1+\omega_2;\omega_1,\omega_2)$ is the second-order NLO conductivity.
In general, the second-order NLO conductivity is given by (second-order) Kubo formula~\cite{Moss1987, Ghahramani1991, Rashkeev2001, Leitsmann2005, Kang2010, Lee2002, Kang2013, Aversa1995, Sipe1993PRB, Xu2021, Sipe2000, Habara2022}:  
\begin{equation}
  \begin{split}
    \sigma^{(2)}_{ijk}&(\omega_1+\omega_2; \omega_1, \omega_2)\\
    &\equiv -\frac{\hbar^2e^2}{2S}\sum_{\bm{k}}(\Omega^{(2)}_{ijk} (\omega_1, \omega_2, \bm{k}) + \Omega^{(2)}_{ikj} (\omega_2, \omega_1, \bm{k}))
  \end{split}
  \label{eq:2ndkubo}
\end{equation}
with
\begin{equation}
 \begin{split}
   \Omega^{(2)}_{ijk}(\omega_1, \omega_2, \bm{k})=&\sum_{nml}\frac{j^{i}_{mn}}{E_{ml}E_{ln}(E_{mn}-\hbar(\omega_1+\omega_2)-i\eta)} \\
& \times\Biggl[\frac{v^j_{nl}v^k_{lm}f_{ml}}{E_{ml}-\hbar\omega_2-i\eta}
    -\frac{v^k_{nl}v^j_{lm}f_{ln}}{E_{ln}-\hbar\omega_2-i\eta}\Biggl]\\
  =&\sum_{nml}\frac{f_{lm}v^k_{lm}}{E_{ml}(E_{ml}-\hbar\omega_2-i\eta)} \\
  & \times\Biggl[\frac{j^i_{mn}v^j_{nl}}{E_{nl}(E_{mn}-\hbar(\omega_1+\omega_2)-i\eta)}\\
    &-\frac{v^j_{mn}j^i_{nl}}{E_{mn}(E_{nl}-\hbar(\omega_1+\omega_2)-i\eta)}\Biggl], 
 \end{split}
 \label{eq:2ndkubo2}
\end{equation}
which describes the contribution of the interband optical transition~\cite{Taghizadeh2017}.
$\Omega^{(2)}_{ijk}(\omega_1, \omega_2, \bm{k})$ is integrand of $\sigma^{(2)}_{ijk}(\omega_1+\omega_2; \omega_1, \omega_2)$ before considering intrinsic permutation symmetry, i.e., $\Omega^{(2)}_{ijk}(\omega_1, \omega_2, \bm{k})=\Omega^{(2)}_{ikj}(\omega_2, \omega_1, \bm{k})$.
Note that $(n, m, l)$ is the band indices including spin degree of freedom.
$v^j_{nl}=\braket{u_{n\bm{k}}|\bm{v}\cdot\bm{e_j}|u_{l\bm{k}}}$, and $\bm{v}$ is the group velocity operator, which written as $\bm{v}=(\hat{v}_x, \hat{v}_y)=\frac{1}{\hbar}(\frac{\partial{\hat{H}}}{\partial{k_x}}, \frac{\partial{\hat{H}}}{\partial{k_y}})$ for 2D materials.
$\bm{e_j}$($\bm{e_k}$) is Jones vector.
For example $\bm{e_x}=(1,0)^T$ for $x$-polarized light, $\bm{e_y}=(0,1)^T$ for $y$-polarized light, $\bm{e_L}=\frac{1}{\sqrt{2}}(1, i)^T$ for LCP light and $\bm{e_R}=\frac{1}{\sqrt{2}}(1, -i)^T$ for RCP light.
$\Ket{u_{n\bm{k}}}$ is the eigenfunction with the eigenenergy $E_{n\bm{k}}$ and $f(E_{n\bm{k}})$ is Fermi-Dirac distribution function.
$E_{ml}$ is the difference between energy levels $E_{m\bm{k}}$ and $E_{l\bm{k}}$, i.e., $E_{ml}\equiv E_{m\bm{k}}-E_{l\bm{k}}$, and $f_{ml}$ is also defined as $f_{ml}\equiv f(E_{m\bm{k}})-f(E_{l\bm{k}})$.
$\omega_1$ and $\omega_2$ are optical angular frequencies for the absorption process.
$\eta$ is infinitesimally small real number and $S$ is the area of 2D
system.
Throughout this paper, $\eta=0.001$ eV is set for the calculation of conductivity.
$j^i_{mn}=\braket{u_{m\bm{k}}|\hat{j}^{(2)}_i|u_{n\bm{k}}}$, where $\hat{j}^{(2)}_i$ is current density operator, and represented as $\hat{j}^{\rm{charge}}_i=\frac{1}{2}\{-e\hat{\sigma}_0\otimes \hat{I}_N, \hat{v}_i\}$ for charge current and $\hat{j}^{\rm{spin}}_i=\frac{1}{2}\{\frac{\hbar}{2}\hat{\sigma}_z\otimes \hat{I}_N, \hat{v}_i\}$ for spin current.
Here, $\hat{I}_N$ is the $N\times N$ identity matrix and in particular $N=3,6,9$ is used for monolayer, bilayer and trilayer NbSe$_2$, respectively.

In Eqs.~(\ref{eq:2ndkubo}) and~(\ref{eq:2ndkubo2}), when $\omega_1+\omega_2 =\omega_3$ and $\omega_1\not=\omega_2\not=\omega_3\not=0$, the process is called as sum frequency generation (SFG)~\cite{Shen1989, Shen2020, Ishiyama2014}.
In particular, if $\omega_1=\omega_2\equiv \omega\not=0$ and $\omega_3=2\omega$, the process becomes SHG.
In addition, when $\omega_1-\omega_2=\omega_3$ and $\omega_1\not=\omega_2\not=\omega_3\not=0$, the process is called as difference frequency generation (DFG)~\cite{Lu2011, Axel2012}, which can express the bulk photovoltaic effect if $\omega_1=-\omega_2\equiv \omega\not=0$.

The NLO conductivities are strongly influenced by the crystal
symmetries~\cite{Boyd2008}. 
For the SHG process in the odd-number-layered NbSe$_2$ where the
inversion symmetry is broken,
the NLO charge and spin conductivities
of SHG process
$\sigma^{\rm{charge}}_{ijk}(2\omega; \omega, \omega)$ and
$\sigma^{\rm{spin}}_{ijk}(2\omega; \omega, \omega)$ 
have following relations:
%
%
\begin{equation}
  \begin{split}
    \sigma^{\rm{charge}}_{yyy}(2\omega; \omega, \omega)&=-\sigma^{\rm{charge}}_{yxx}(2\omega; \omega, \omega) \\
    & =-\sigma^{\rm{charge}}_{xyx}(2\omega; \omega, \omega)=-\sigma^{\rm{charge}}_{xxy}(2\omega; \omega, \omega)
  \end{split}
  \label{eq:shgcharge}
\end{equation}
and 
\begin{equation}
  \begin{split}
    \sigma^{\rm{spin}}_{xxx}(2\omega; \omega, \omega)&=-\sigma^{\rm{spin}}_{xyy}(2\omega; \omega, \omega) \\
    & =-\sigma^{\rm{spin}}_{yxy}(2\omega; \omega, \omega)=-\sigma^{\rm{spin}}_{yyx}(2\omega; \omega, \omega).
  \end{split}
  \label{eq:shgspin}
\end{equation}
The other elements are absent.
On the other hand, in even-number-layered NbSe$_2$, the NLO current is always absent owing to the inversion symmetry.
These results are summarized in Table~\ref{tab:table1}.
In addtion, the another derivation of these relations using crystal symmetry
operation is presented in Appendix A.

\section{NLO conductivity under irradiation of BCL}\label{sec:level4}
\begin{figure*}[t]
  \begin{center}
    \includegraphics[width=1.0\textwidth]{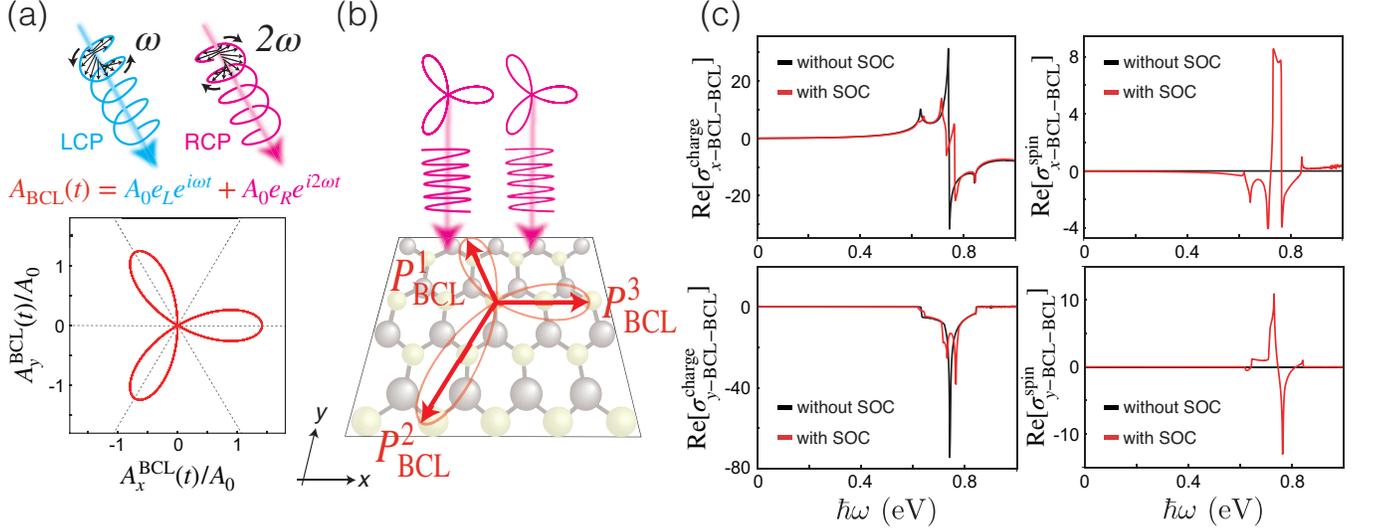}
    \caption{(a) Trajectory of $3$-leaf BCL ($n_1:n_2=1:2$, $\theta=0$).
      (b) Schematic of three induced polarizations $P_{\rm{BCL}}^1$, $P_{\rm{BCL}}^2$ and $P_{\rm{BCL}}^3$ by irradiating $3$-leaf BCL.
      (c) Real parts of second-order NLO conductivities $\sigma^{\rm{charge}}_{x-\rm{BCL}-\rm{BCL}}(n_1\omega, n_2\omega)$, $\sigma^{\rm{charge}}_{y-\rm{BCL}-\rm{BCL}}(n_1\omega, n_2\omega)$, $\sigma^{\rm{spin}}_{x-\rm{BCL}-\rm{BCL}}(n_1\omega, n_2\omega)$ and $\sigma^{\rm{spin}}_{y-\rm{BCL}-\rm{BCL}}(n_1\omega, n_2\omega)$ for monolayer NbSe$_2$ under irradiation of BCL with $3$-leaf.
      Black and red lines indicate the NLO conductivities without and with SOC.
      The units of $\sigma^{\rm{charge}}_{i-\rm{BCL}-\rm{BCL}}(n_1\omega, n_2\omega)$ and $\sigma^{\rm{spin}}_{i-\rm{BCL}-\rm{BCL}}(n_1\omega, n_2\omega)$ are $e^3/\hbar$ and $e^2$, respectively.
    }
    \label{fig:2}
  \end{center}
\end{figure*}
BCL can artificially control the symmetry of electronic system~\cite{Ikeda2022, Dorney2021, Trevisan2022, JimNat2020, Odzak2015, Mauger2016, Chen2018, Gazibegovic2018, Pisanty2014}.
BCL is the superposition of LCP and RCP light with different
angular frequencies $n_1\omega$ and $n_2\omega$, where $n_1$ and $n_2$
are different integers.
The vector potential for BCL can be formulated as
\begin{equation}
  \begin{split}
    A(t) & =A_Le^{in_1\omega t}+A_Re^{in_2\omega t-i\theta}+{\rm c.c.} \\
    & = \frac{1}{\sqrt{2}}(A_x+iA_y)e^{in_1\omega t}+\frac{1}{\sqrt{2}}(A_x-iA_y)e^{in_2\omega t-i\theta}+{\rm c.c.},
  \end{split}
  \label{eq:bcl}
\end{equation}
where $A_{L(R)}$ is the amplitude of the LCP(RCP) light.
$\theta$ is the phase difference between LCP and RCP light.
In general, the trajectory of BCL is a rose curve with $(n_1+n_2)/\textrm{gcd}(n_1, n_2)$-fold rotation symmetry.
Here, gcd($n_1, n_2$) means the great common divisor of two integers
$n_1, n_2$. 
For example, there are $3$-leaf BCL with $3$-fold rotation symmetry $C_3$, $4$-leaf BCL with $4$-fold rotation symmetry $C_4$ and $5$-leaf BCL with $5$-fold rotation symmetry $C_5$.


The induced current density $J^{\rm{BCL}}_{i}$ under the
irradiation of BCL, i.e., $E_{\rm{BCL}}(t)=E_Le^{in_1\omega t}+E_Re^{in_2\omega t}+{\rm c.c.}=\sum_{a=L, R}\sum_{n_a}E_ae^{in_a\omega t}$ with $(n_L, n_R)\equiv(n_1, n_2)$, can be written as
\begin{equation}
  \begin{split}
    J^{\rm{BCL}}_i = \sum_{ab}\sum_{n_a n_b}\sigma^{(2)}_{iab}((n_a+n_b)\omega; n_a\omega, n_b\omega)E_aE_b.
    \nonumber
  \end{split}
\end{equation}
Furthermore, the NLO conductivity for BCL irradiation becomes
\begin{equation}
  \begin{split}
    &\sigma^{(2)}_{i-\rm{BCL}-\rm{BCL}}(n_1\omega, n_2\omega)\\
    &\quad= \sum_{ab}\sigma^{(2)}_{iab}((n_a+n_b)\omega; n_a\omega, n_b\omega)\\
    &\quad= \sigma^{(2)}_{iLL}(2n_1\omega; n_1\omega, n_1\omega)+\sigma^{(2)}_{iLR}((n_1+n_2)\omega; n_1\omega, n_2\omega)\\
    &\qquad+\sigma^{(2)}_{iRL}((n_2+n_1)\omega; n_2\omega, n_1\omega)+\sigma^{(2)}_{iRR}(2n_2\omega; n_2\omega, n_2\omega),
    \label{eq:curdensbcl}
  \end{split}
\end{equation}
where the subscript BCL and L (R) mean the injection of BCL and LCP (RCP),
respectively.
The NLO conductivities are numerically calculated by using Eqs.~(\ref{eq:2ndkubo}) and~(\ref{eq:2ndkubo2}).

\begin{figure}[h]
  \begin{center}
    \includegraphics[width=0.5\textwidth]{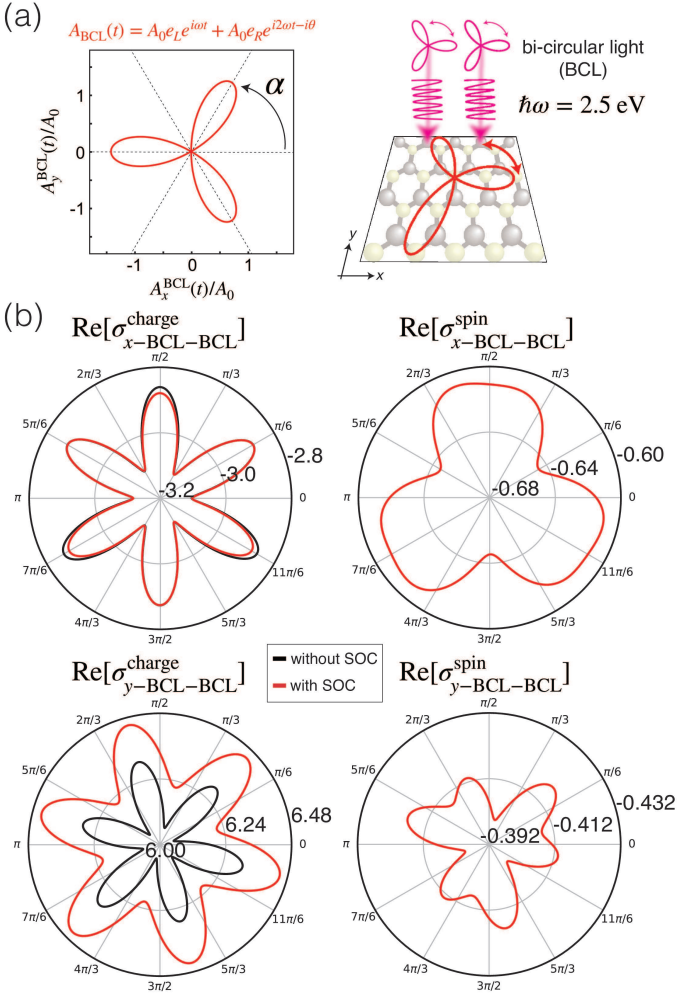}
    \caption{(a) Trajectory of $3$-leaf BCL with rotational angle $\alpha$ ($n_1:n_2=1:2$, $\alpha\not=0$).
      (b) $\alpha$-dependence on real parts of second-order NLO charge and spin conductivities for monolayer NbSe$_2$ under irradiation of $3$-leaf BCL of $\hbar\omega=2.5$ eV.
      Black and red lines indicate $\sigma^{(2)}_{i-\rm{BCL}-\rm{BCL}}(n_1\omega, n_2\omega)$ without and with SOC.
      The units of $\sigma^{\rm{charge}}_{i-\rm{BCL}-\rm{BCL}}(n_1\omega, n_2\omega)$ and $\sigma^{\rm{spin}}_{i-\rm{BCL}-\rm{BCL}}(n_1\omega, n_2\omega)$ are $e^3/\hbar$ and $e^2$, respectively.}
    \label{fig:3}
  \end{center}
\end{figure}
Figure~\ref{fig:2} (a) shows the trajectory of incident BCL with
$3$-leaf ($n_1:n_2=1:2$, $\theta=0$).
The $3$-leaf BCL can break the mirror symmetry $M_x$ and then induce three polarizations $P_{\rm{BCL}}^1$, $P_{\rm{BCL}}^2$ and $P_{\rm{BCL}}^3$ along the directions of three leaves of $3$-leaf BCL (see Fig.~\ref{fig:2} (b)).
Figure~\ref{fig:2} (c) shows the $\omega$ dependence of the real parts
of 
second-order NLO conductivities $\sigma^{(2)}_{i-\rm{BCL}-\rm{BCL}}(n_1\omega, n_2\omega)$
for monolayer NbSe$_2$ under irradiation of $3$-leaf BCL.
In previous work, we have shown that under LP light irradiation, the charge and spin Hall current can be generated only either in $y$- or $x$-direction for monolayer NbSe$_2$~\cite{Habara2022}. 
However, since the $3$-leaf BCL breaks $M_x$ symmetry, the current can
be generated in both $x$- and $y$-directions. 

Figure~\ref{fig:3} (a) shows the trajectory of $3$-leaf BCL with
rotational angle $\alpha$ ($n_1:n_2=1:2$, $\alpha\not=0$). 
Here, $\alpha$ has a following relation with $\theta$: 
\begin{equation}
  \alpha = -\frac{n_1}{n_1+n_2}\theta, 
  \label{eq:alpha}
\end{equation}
which is defined as rotational angle from $x$-axis.
Figure~\ref{fig:3} (b) shows $\alpha$-dependence on real parts of
second-order NLO conductivities
$\sigma^{(2)}_{i-\rm{BCL}-\rm{BCL}}(n_1\omega, n_2\omega)$
for monolayer NbSe$_2$ under irradiation of $3$-leaf BCL of
$\hbar\omega=2.5$ eV. 
The charge current is generated along the three different axes of the
induced polarizations by incident BCL of $3$-leaf, resulting in the
appearance of six equivalent peaks for $\alpha$
($0\leq\alpha\textless2\pi$). 
On the other hand, the magnitude of spin current has only three
 equivalent peaks.
As shown in Supplementary material, the NLO conductivities for up- and
 down-spin states have three equivalent peaks, but point in opposite
 directions to each other. 
In general, the number of generated directions of charge and spin current are twice as different.

In Appendix~A, we derive the selection rules of second-order NLO
conductivities of NbSe$_2$ under irradiation of generic BCL.
In Appendix~B, we provide the second-order NLO conductivities of monolayer NbSe$_2$ for the
irradiation of BCL with $4$-, $5$-, $6$- and $7$-leaf.
As shown in Fig.~\ref{fig:8} of Appendix~B, it is found that 
the peaks of NLO conductivities in the range of $0<\hbar\omega<1$
shift lower with increase of leaves of BCL. 
These shifts are originated from the factor of $(n_1+n_2)$ in the 
denominator of Eq.~(\ref{eq:2ndkubo2}).


If $n_1:n_2=1:1$ with $\theta=0$, the trajectory of incident light is
identical to $x$-polarized light, i.e.,
\begin{equation}
  \begin{split}
    A_{\rm{BCL}}(t)=A_Le^{i\omega t}+A_Re^{i\omega t}+{\rm c.c.}=\sqrt{2}A_xe^{i\omega t}+{\rm c.c.}.
  \end{split}
\end{equation}
Therefore, this case reproduces the results of SHG process~\cite{Habara2022}. 
%
The details are shown in Supplementary material~\cite{supplementary}. 
\begin{figure*}[t]
  \begin{center}
    \includegraphics[width=0.90\textwidth]{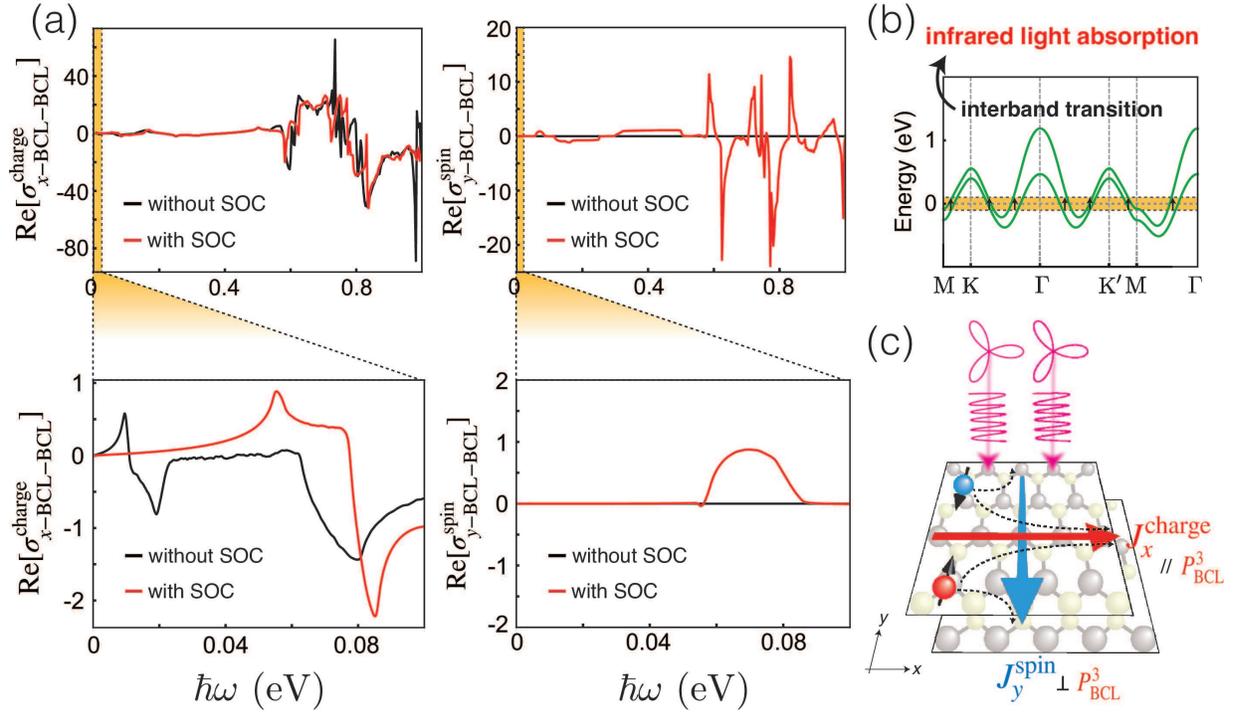}
    \caption{(a) Real parts of second-order NLO conductivities $\sigma^{\rm{charge}}_{x-\rm{BCL}-\rm{BCL}}(n_1\omega, n_2\omega)$ and $\sigma^{\rm{spin}}_{y-\rm{BCL}-\rm{BCL}}(n_1\omega, n_2\omega)$ for bilayer NbSe$_2$ under irradiation of BCL with $3$-leaf.
      In an energy range of $0<\hbar\omega<0.1$, the charge and spin conductivities have peaks.
      (b) Its peaks correspond to infrared light absorption for optical interband transition in energy band of bilayer NbSe$_2$.
      (c) Schematics of generated charge and spin current by irradiating BCL with $3$-leaf. 
      Black and red lines indicate $\sigma^{(2)}_{i-\rm{BCL}-\rm{BCL}}(n_1\omega, n_2\omega)$ without and with SOC.
      The units of $\sigma^{\rm{charge}}_{i-\rm{BCL}-\rm{BCL}}(n_1\omega, n_2\omega)$ and $\sigma^{\rm{spin}}_{i-\rm{BCL}-\rm{BCL}}(n_1\omega, n_2\omega)$ are $e^3/\hbar$ and $e^2$, respectively.}
    \label{fig:4}
  \end{center}
\end{figure*}

In even-number-layered NbSe$_2$, the second-order NLO current is always absent because of the inversion
symmetry~\cite{Habara2022}.  
However, the irradiation of BCL with odd-number of leaf or even-number
of leaf with $\alpha\neq 0$ can break the inversion symmetry, resulting in the
generation of charge and spin current.
As shown in Appendix~C, for irradiation of RCP and LCP light with
different optical angular frequencies, the NLO current is generated,
i.e., no cancellation of current even in the presence of spatial
inversion symmetry.
Figure~\ref{fig:4} (a) shows the real parts of 
$\sigma^{(2)}_{i-\rm{BCL}-\rm{BCL}}(n_1\omega,n_2\omega)$ for 
bilayer NbSe$_2$ under irradiation of $3$-leaf BCL.   
%
Owing to the broken mirror symmetry $M_x$, 
the NLO charge and spin current can be generated only either in $x$- or
$y$-direction for bilayer NbSe$_2$. 
Thus, the irradiation of BCL can generate the current even
in even-number-layered NbSe$_2$.

In further, we should notice that in an energy range of
$0<\hbar\omega<0.1$, $\mathrm{Re}[\sigma^{\rm{charge}}_{x-\rm{BCL}-\rm{BCL}}]$ and $\mathrm{Re}[\sigma^{\rm{spin}}_{y-\rm{BCL}-\rm{BCL}}]$ have the peaks for bilayer NbSe$_2$. 
These peaks are attributed to interband
optical transition between
highest valence and lowest conduction bands of energy bands shown in Fig.~\ref{fig:4} (b), which cause infrared light absorption.
On the other hand, the infrared light absorption is absent in
semiconductor TMDC such as MoS$_2$ because of the large band gap between
valence and conduction bands ($\Delta E\approx2.0$ eV).

Figure~\ref{fig:4} (c) schematically summarizes the induced
charge and spin current directions in bilayer NbSe$_2$ under irradiation of $3$-leaf BCL.
Since BCL breaks the inversion symmetry of electronic system, the NLO current is generated even in bilayer NbSe$_2$.
Note that the charge current is induced along $P^3_{\rm{BCL}}$ in Fig.~\ref{fig:2} (b), but the
spin current is induced in perpendicular to the charge current.

In Appendix~D, we provide the second-order NLO conductivities for 
trilayer NbSe$_2$ under BCL irradiation.
The second-order NLO conductivities of trilayer NbSe$_2$
consist of three optical
transition processes: (i) intralayer optical transition of monolayer
NbSe$_2$, (ii) intra- and interlayer optical transition of bilayer
NbSe$_2$ and (iii) interlayer optical transitions between monolayer and
bilayer NbSe$_2$. 
Under irradiation of LP light, the process (ii) is absent.
However, BCL of $3$-leaf makes 
all three processes finite.
%

\section{Shift- and injection-current conductivities}\label{sec:level5}
\begin{figure*}[t]
  \begin{center}
    \includegraphics[width=0.95\textwidth]{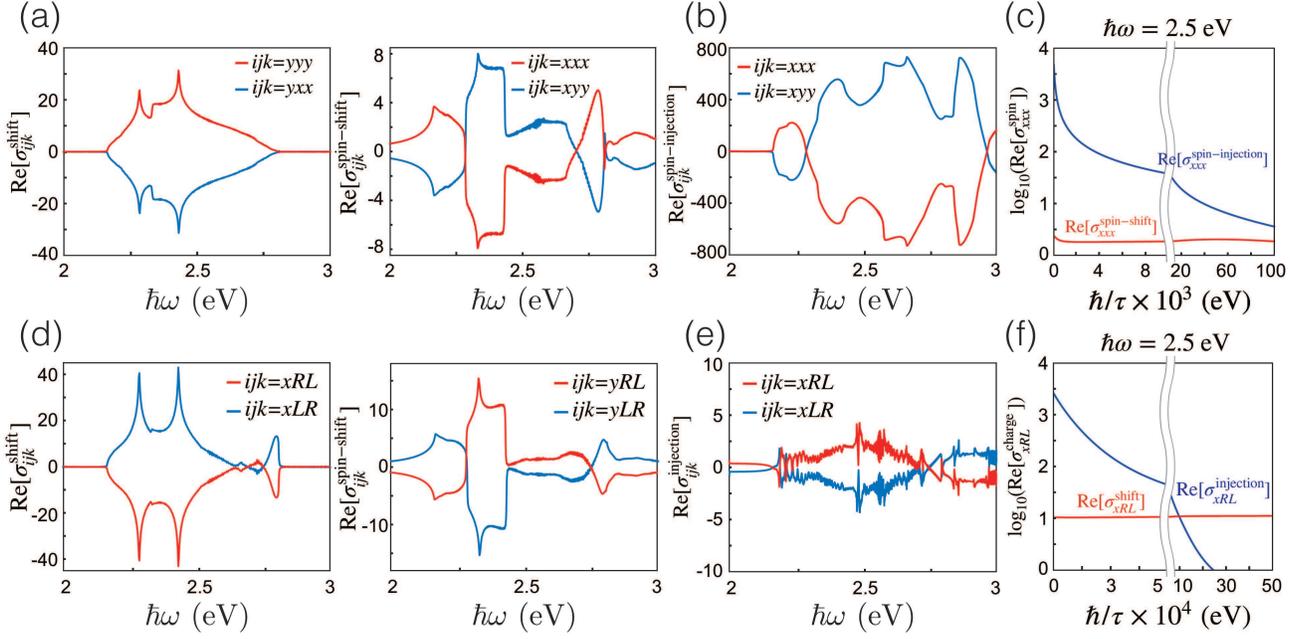}
    \caption{Real parts of charge and spin (a, d) shift- and (b, e) injection-current conductivities for monolayer NbSe$_2$ under LP and CP light irradiation, respectively.
      (c, f) $\tau$-dependences of shift (red) and injection (blue) current conductivities for linearly $x$-polarized and RCP light of $\hbar\omega=2.5$ eV.
      Here, plots of the NLO conductivities are shown in logarithmic scale.
      The units of NLO charge and spin conductivities are $e^3/\hbar$ and $e^2$, respectively.}
    \label{fig:5}
  \end{center}
\end{figure*}
DFG with $\omega_1=-\omega_2\equiv\omega\not=0$ induces the
DC photocurrent, i.e., bulk photovoltaic effect. 
In Eq.~(\ref{eq:j2}), the DC photocurrent $J^{(2)}_i$ can be obtained as 
\begin{equation}
  \begin{split}
    J^{(2)}_i = &\sum_{jk}\{\sigma^{(2)}_{ijk}(0;\omega, -\omega)E_j(\omega)E_k(-\omega) \\
    & +\sigma^{(2)}_{ijk}(0;-\omega, \omega)E_j(-\omega)E_k(\omega)\}\\
    =& \sum_{jk}\{\sigma^{(2)}_{ijk}(0; \omega, -\omega)E_j(\omega)E_k(-\omega)+{\rm c.c.}\}\\
    =& \sum_{jk}2\sigma^{(2)}_{ijk}(0; \omega, -\omega)E_j(\omega)E_k(-\omega),
  \end{split}
  \label{eq:acanddcomega}
\end{equation}
where the DC photoconductivity based on Eqs.~(\ref{eq:2ndkubo}) and~(\ref{eq:2ndkubo2}) follows the identities $\sigma^{(2)}_{ijk}(0; -\omega, \omega)=-\sigma^{(2)}_{ijk}(0; \omega, -\omega)=[\sigma^{(2)}_{ijk}(0; \omega, -\omega)]^*$.
%
%
%
The numerical results are shown in Supplementary material~\cite{supplementary}, which strongly depends on the crystal symmetry, i.e., generation of current for odd-number-layered NbSe$_2$, but absence for even-number-layered NbSe$_2$.
The bulk photovoltaic effect includes the generation of shift and injection current.
In this section, we derive the shift/injection-current conductivities, and find that these conductivities depend on the crystal symmetry and polarization of incident light for monolayer NbSe$_2$.

%

\subsection{Shift and injection current under LP light irradiation}
The DC photocurrent can be separated into two different optically induced DC current, i.e., (i) shift current and (ii) injection current.
(i) Shift current is induced polarization current owing to the shift of electron position by irradiating light~\cite{Xu2021}.
The induced polarization of electrons $P_i$ is written as
\begin{equation}
  P_i=e\int_{\rm{BZ}}\frac{d\bm{k}}{(2\pi)^3}\sum_mf_m\xi^i_{mm}, 
\end{equation}
where $\xi^i_{mm}$ is the Berry connection of $m$-th band and
$\xi^i_{mm}=i\braket{u_{m\bm{k}}|\frac{\partial}{\partial{k_i}}|u_{m\bm{k}}}$~\cite{Sipe2000}. 
Berry connection is also interpreted as the Wannier center of wave
function~\cite{King1993, Resta1994, Marzari2012}. 
Thus, the origin of the shift current is the change of the polarization
$P_i$, i.e., the Berry connection difference between valence and
conduction bands $\xi^i_{mm}-\xi^i_{ll}$ by photo-excitation.  
Note that $l (m)$ is the unoccupied (occupied) band index.
(ii) Injection current is induced net current for the distribution of asymmetric electron and hole velocities $\Delta^i_{lm}=v^i_{ll}-v^i_{mm}$ owing to the population imbalance by photo-excitation in the momentum space~\cite{Xu2021}.

Using a sum rule for the generalized derivative $r^j_{ml;i}$ of position $r^j_{ml}=\braket{u_{m\bm{k}}|\bm{r}\cdot\bm{e_j}|u_{l\bm{k}}}$ in the DC photoconductivity $\sigma^{(2)}_{ijk}(0; \omega, -\omega)$~\cite{Xu2021, Sipe2000, Fei2020, Bieler2006, Fregoso2019}, the DC photoconductivity can be separated into shift/injection-current conductivities. 
Here, $r^j_{ml;i}$ is given as
\begin{equation}
  \begin{split}
    r^j_{ml;i}&=\frac{\partial{r^j_{ml}}}{\partial{k_i}}-i[\xi^i_{mm}-\xi^i_{ll}]r^j_{ml}\\
    &=-iR^i_{ml}(\bm{k})r^j_{ml},  
  \end{split}
  \label{eq:sumrulexi}
\end{equation}
which depends on the difference of Berry connections, i.e., contribution for shift current.
Note that $R^i_{ml}(\bm{k})$ is shift vector and given as
\begin{equation}
  R^i_{ml}(\bm{k})=\frac{\partial{\phi^j_{ml}}}{\partial{k_i}}+(\xi^i_{mm}-\xi^i_{ll}), 
\end{equation}
where $\phi^j_{ml}$ is phase of group velocity $v^j_{ml}=\braket{u_{m\bm{k}}|\bm{v}\cdot\bm{e_j}|u_{l\bm{k}}}=|v^j_{ml}|e^{-i\phi^j_{ml}}$.
For the numerical calculation, it is better to rewrite the generalized derivative $r^j_{ml;i}$ as follow, 
\begin{equation}
  \begin{split}
    r^j_{ml;i}=\frac{r^i_{ml}\Delta^j_{lm}+r^j_{ml}\Delta^i_{lm}}{\omega_{ml}} 
    -i\hbar^2\sum_{n\not=m,l}\Biggl(\frac{v^i_{mn}v^j_{nl}}{E_{ml}E_{mn}}-\frac{v^j_{mn}v^i_{nl}}{E_{ml}E_{nl}}\Biggl).
  \end{split}
  \label{eq:sumrule}
\end{equation}
Using the relation of $r^i_{ml}$ and $v^i_{ml}$ for interband optical transition, i.e.,
\begin{align}
  r^i_{ml} \ =\
  \begin{cases}
    \frac{v^i_{ml}}{i\omega_{ml}} & (m\not=l) \\
    0 & (m=l)
  \end{cases}, 
\end{align}
Eq.~(\ref{eq:sumrule}) is rewritten as the expression with velocity.
Thus, $r^j_{ml;i}$ becomes
\begin{equation}
  \begin{split}
    r^j_{ml;i}&=-\hbar R^i_{ml}(\bm{k})\frac{v^j_{ml}}{E_{ml}}\\
    &= -\frac{2i\hbar^2}{E_{ml}}\Bigg[\frac{v^i_{ml}(v^j_{ll}-v^j_{mm})}{E_{ml}}+\frac{v^j_{ml}(v^i_{ll}-v^i_{mm})}{E_{ml}} \\
      & +\sum_{n}\Bigg(\frac{v^i_{mn}v^j_{nl}}{E_{mn}}-\frac{v^j_{mn}v^i_{nl}}{E_{nl}}\Bigg)\Bigg], 
  \end{split}
  \label{eq:numericalR}
\end{equation}
where $v^i_{ml}$ is $\braket{u_{m\bm{k}}|\hat{j}^{\rm{charge}}_{i}|u_{l\bm{k}}}=\frac{1}{2}\braket{u_{m\bm{k}}|\{-e \hat{\sigma}_0\otimes\hat{I}_N, \hat{v}_i\}|u_{l\bm{k}}}$ for charge current and $\braket{u_{m\bm{k}}|\hat{j}^{\rm{spin}}_{i}|u_{l\bm{k}}}=\frac{1}{2}\braket{u_{m\bm{k}}|\{\frac{\hbar}{2}\hat{\sigma}_z\otimes\hat{I}_N, \hat{v}_i\}|u_{l\bm{k}}}$ for spin current.
Thus, by substituting the second-term of Eq.~(\ref{eq:numericalR}) in the summation part of Eq.~(\ref{eq:2ndkubo2}) with $\omega_1=-\omega_2\equiv\omega\not=0$, the DC photoconductivity becomes $\sigma^{(2)}_{ijk}(0; \omega, -\omega)=\sigma^{\rm{shift}}_{ijk}(0; \omega, -\omega)+\sigma^{\rm{injection}}_{ijk}(0; \omega, -\omega)$.
Here, $\sigma^{\rm{shift}}_{ijk}(0; \omega, -\omega)$ is shift-current and $\sigma^{\rm{injection}}_{ijk}(0; \omega, -\omega)$ is injection-current conductivities.

For LP light irradiation, $\sigma^{\rm{shift}}_{ijk}(0; \omega, -\omega)$ and $\sigma^{\rm{injection}}_{ijk}(0; \omega, -\omega)$ are derived as~\cite{Fei2020}
\begin{equation}
  \sigma^{\rm{shift}}_{ijk}(0; \omega, -\omega)=-\frac{i\hbar e^3}{S}\sum_{\bm{k}}\sum_{mn}f_{nm}\frac{\alpha^{jk}_{mn}(\bm{k})R^i_{nm}(\bm{k})}{E_{mn}^2(E_{mn}-\hbar\omega-i\eta)}
  \label{eq:shiftcur}
\end{equation}
with transition intensity 
\begin{equation}
  \begin{split}
    \alpha^{jk}_{mn}(\bm{k})=\frac{1}{2}(v^j_{mn}(\bm{k})v^k_{nm}(\bm{k})+v^k_{mn}(\bm{k})v^j_{nm}(\bm{k})), 
    \label{eq:RphiXi}
  \end{split}
\end{equation}
and
\begin{equation}
  \sigma^{\rm{injection}}_{ijk}(0; \omega, -\omega)=\tau\frac{i\hbar e^3}{S}\sum_{\bm{k}}\sum_{mn}f_{nm}\frac{\alpha^{jk}_{mn}(\bm{k})\Delta^i_{mn}(\bm{k})}{E_{mn}^2(E_{mn}-\hbar\omega-i\eta)}, 
  \label{eq:injectcur}
\end{equation}
where the dummy variables ($m\rightarrow n, l\rightarrow m$) are interchanged and $\tau=\hbar/\eta$. 
For spin current, the superscripts of these conductivities are rewritten as ``$\rm{spin-shift}$'' and ``$\rm{spin-injection}$''.
It should be noted that the spin-dependent position and velocity difference for the direction of generated current are $r^j_{ml;i(\rm{spin})}$ and $\Delta^{i-\rm{spin}}_{lm}$. 
According to Ref.~\cite{Xu2021}, since these approaches of length~\cite{Sipe2000} and velocity gauges~\cite{Fei2020} should be equivalent, we use the expression of velocity gauge for our calculation.

Since the shift current strongly depends on the crystal symmetry, 
the charge and spin shift current can be generated in odd-number-layered
NbSe$_2$, but absent in even-number-layered NbSe$_2$. 
The finite charge and spin shift-current conductivities can be obtained as
\begin{equation}
  \begin{split}
    \sigma^{\rm{shift}}_{yyy}(0; \omega, -\omega) & =-\sigma^{\rm{shift}}_{yxx}(0; \omega, -\omega) \\
    & =-\sigma^{\rm{shift}}_{xyx}(0; \omega, -\omega)=-\sigma^{\rm{shift}}_{xxy}(0; \omega, -\omega)
  \end{split}
  \label{eq:shiftrelation}
\end{equation}
and 
\begin{equation}
  \begin{split}
    \sigma^{\rm{spin-shift}}_{xxx}&(0; \omega, -\omega)=-\sigma^{\rm{spin-shift}}_{xyy}(0; \omega, -\omega) \\
    &=-\sigma^{\rm{spin-shift}}_{yxy}(0; \omega, -\omega)=-\sigma^{\rm{spin-shift}}_{yyx}(0; \omega, -\omega), 
  \end{split}
  \label{eq:spinshiftrelation}
\end{equation}
respectively.
The other elements are absent for monolayer NbSe$_2$.
These relations are equivalent to the results of SHG process~\cite{Habara2022}.

Figure~\ref{fig:5} (a) shows the charge and spin shift-current
conductivities for monolayer NbSe$_2$ under
irradiation of LP light. 
Because of the broken inversion symmetry, the charge and spin shift current can be generated in $y$- and $x$-directions for linearly $x$-polarized light, respectively. 
On the other hand, for linearly $y$-polarized light, the charge and spin shift current can be generated in $y$- and $x$-directions, respectively.
We should note that the generated spin shift current is perpendicular
to the charge current. 

%
%
The appearance of charge (spin) shift current along $y$ ($x$)-direction can also be
understood by inspecting $\alpha^{jk}_{mn}(\bm{k})$ with $jk=xx,yy$
and $R^i_{nm}(\bm{k})$, which are contained 
in the intergrand of Eq.~(\ref{eq:shiftcur}).
When the product $\alpha^{jk}_{mn}(\bm{k})R^i_{nm}(\bm{k})$ with $jk=xx,yy$
is even with respect to $\pi$-rotation, we obtain the finite charge (spin) shift current along $y$ ($x$)-direction.
The contour plots of $\alpha^{jk}_{mn}(\bm{k})$ and $R^i_{nm}(\bm{k})$
can be found in Fig.~\ref{fig:11} in Appendix~E.
Since $\alpha^{jk}_{mn}(\bm{k})$ with $jk=xx,yy$ is even
for $\pi$-rotation, the shift current is absent for $R^i_{mm}(\bm{k})$ with odd parity.

\begin{table*}
  \caption{\label{tab:table2}Shift- and injection-current conductivities for monolayer NbSe$_2$ under irradiation of LP and CP light.}
\begin{ruledtabular}
\begin{tabular}{ccccc}
 &$\sigma^{\rm{shift}}_{ijk}(0; \omega, -\omega)$&$\sigma^{\rm{spin-shift}}_{ijk}(0; \omega, -\omega)$&$\sigma^{\rm{injection}}_{ijk}(0; \omega, -\omega)$&$\sigma^{\rm{spin-injection}}_{ijk}(0; \omega, -\omega)$ \\ \hline
  $ijk=xxx$&$0$&$\approx10^0$&$0$&$\approx10^2$ ($\eta=0.001$ eV) \\
  $ijk=xyy,yxy,yyx$&$0$&$\approx-10^0$&$0$&$\approx-10^2$ ($\eta=0.001$ eV) \\
  $ijk=yyy$&$\approx10^1$&$0$&$0$&$0$ \\
  $ijk=yxx,xyx,xxy$&$\approx-10^1$&$0$&$0$&$0$ \\
  $ijk=xRL$&$\approx10^1$&$0$&$\approx10^1$ ($\eta=0.001$ eV)&$0$ \\
  $ijk=yRL$&$0$&$\approx10^1$&$0$&$0$ \\
  $ijk=xLR$&$\approx-10^1$&$0$&$\approx-10^1$ ($\eta=0.001$ eV)&$0$ \\
  $ijk=yLR$&$0$&$\approx-10^1$&$0$&$0$ \\
\end{tabular}
\end{ruledtabular}
\end{table*}
On the other hand, the finite elements of injection-current conductivities have a following relation:
\begin{equation}
  \begin{split}
    &\sigma^{\rm{spin-injection}}_{xxx}(0; \omega, -\omega)=-\sigma^{\rm{spin-injection}}_{xyy}(0; \omega, -\omega) \\
    &=-\sigma^{\rm{spin-injection}}_{yxy}(0; \omega, -\omega)=-\sigma^{\rm{spin-injection}}_{yyx}(0; \omega, -\omega), 
  \end{split}
  \label{eq:spininjectionrelation}
\end{equation}
however, the other elements are absent for monolayer NbSe$_2$.

Figure~\ref{fig:5} (b) shows the spin injection-current conductivities for monolayer NbSe$_2$ under LP light irradiation.
Owing to the broken inversion symmetry, the spin injection current can be generated in $x$-direction for $x$- and $y$-polarized light.
As shown in Fig.~\ref{fig:5} (c), the magnitude of the generated spin injection current is larger than that of spin shift current owing to $\tau$-dependence of injection current.
Here, the plots of conductivities are shown in logarithmic scale.
Thus, if $\tau$ increases, the magnitude of injection current becomes larger, but that of shift current is robust, i.e., topologically protected.

We also find that the parity of 
$\alpha^{jk}_{mn}(\bm{k})\Delta^i_{mn}(\bm{k})$ in the integrand of
Eq.~(\ref{eq:injectcur}) is involved in the generation of injection
current. 
As shown in Fig.~\ref{fig:11} in Appendix~E, $\alpha^{jk}_{mn}(\bm{k})$
with $jk=xx,yy$ and $\Delta^{x-\rm{spin}}_{mn}(\bm{k})$ are even for $\pi$-rotation. 
Thus, the spin injection current
is generated in $x$-direction, because 
$\alpha^{jk}_{mn}(\bm{k})\Delta^i_{mn}(\bm{k})$ with $jk=xx,yy$ is
even with respect to $\pi$-rotation.

\subsection{Shift and injection current under CP light irradiation}
Similarly, for irradiation of CP light, the shift/injection-current conductivities are given as
\begin{equation}
  \begin{split}
    \sigma^{\rm{shift}}_{iRL}(0; \omega, -\omega) = \frac{e^3}{\hbar S}\sum_{\bm{k}}\sum_{mn}f_{nm}\frac{\Omega^{xy}_{mn}(\bm{k})-\Omega^{yx}_{mn}(\bm{k})}{E_{mn}-\hbar\omega-i\eta}R^i_{nm}(\bm{k})
  \end{split}
  \label{eq:RLshiftcurrent}
\end{equation}
and
\begin{equation}
  \begin{split}
    \sigma^{\rm{injection}}_{iRL}(0; \omega, -\omega) = \frac{\tau e^3}{\hbar S}\sum_{\bm{k}}\sum_{mn}f_{nm}\frac{\Omega^{xy}_{mn}(\bm{k}) - \Omega^{yx}_{mn}(\bm{k})}{E_{mn}-\hbar\omega-i\eta}\Delta^i_{mn}(\bm{k})
  \end{split}
  \label{eq:RLinjectioncurrent}
\end{equation}
with Berry curvature 
\begin{equation}
  \begin{split}
    \Omega^{xy}_{mn}(\bm{k}) = -\Omega^{yx}_{mn}(\bm{k}) = -\hbar^2\frac{\mathrm{Im}(v^{x}_{mn}(\bm{k})v^{y}_{nm}(\bm{k}))}{E_{mn}^2}.
  \end{split}
\end{equation}
Note that the subscript $RL$ means irradiation of RCP light. 
If these conductivities include the spin current operator, Eqs.~(\ref{eq:RLshiftcurrent}) and~(\ref{eq:RLinjectioncurrent}) are rewritten as spin-dependent shift- and injection-current conductivities. 
Here, $R^i_{nm}(\bm{k})$ and $\Delta^i_{mn}(\bm{k})$ includes the charge and spin operators. 

For RCP light, the finite elements of charge and spin
shift-current conductivities are $\sigma^{\rm{shift}}_{xRL}(0; \omega, -\omega)$ and $\sigma^{\rm{spin-shift}}_{yRL}(0; \omega, -\omega)$.
For LCP light, the charge and spin shift-current conductivities have opposite sign to those of RCP light, i.e., 
\begin{equation}
  \begin{split}
    \sigma^{\rm{shift}}_{xLR}(0; \omega, -\omega) & =-\sigma^{\rm{shift}}_{xRL}(0; \omega, -\omega), \\ 
    \sigma^{\rm{spin-shift}}_{yLR}(0; \omega, -\omega) & =-\sigma^{\rm{spin-shift}}_{yRL}(0; \omega, -\omega).
  \end{split}
\end{equation}
The other elements are absent for monolayer NbSe$_2$.

Figure~\ref{fig:5} (d) shows the charge and spin shift-current
conductivities for monolayer NbSe$_2$ under irradiation of CP light. 
Because of the broken inversion symmetry, the charge shift current can be generated in $x$-direction for CP light.
On the other hand, the spin shift current can be generated in $y$-direction for CP light.
Comparing the results of CP light with those of LP light, the directions
of generated charge and spin shift current 
are switched for monolayer NbSe$_2$: the charge shift current is
generated in $y$-direction for LP light, but in $x$-direction for CP
light. For spin shift current, the irradiation of LP and CP light causes
the generation in $x$- and $y$-directions, respectively. 
Thus, the direction of generated shift current can be switched if LP light is altered to CP light.

For CP light, the charge injection-current conductivities $\sigma^{\rm{injection}}_{xRL}(0; \omega, -\omega)$ and $\sigma^{\rm{injection}}_{xLR}(0; \omega, -\omega)$ are finite for monolayer NbSe$_2$, in further the relation of these finite conductivities is given as 
\begin{equation}
  \begin{split}
    \sigma^{\rm{injection}}_{xRL}(0; \omega, -\omega)=-\sigma^{\rm{injection}}_{xLR}(0; \omega, -\omega).
  \end{split}
\end{equation}
The other elements are absent for monolayer NbSe$_2$.

Figure~\ref{fig:5} (e) shows the charge injection-current conductivities for monolayer NbSe$_2$ under irradiation of CP light.
Owing to the broken inversion symmetry, the charge injection current can be generated in $x$-direction for CP light.
As shown in Fig.~\ref{fig:5} (f), the magnitude of injection-current
conductivity becomes larger with the increase of $\tau$, but that of
shift-current conductivity is robust owing to its topological properties.
It is found that the generated injection current for CP light irradiation has stronger $\tau$-dependence than that of LP light.


$\Omega^{xy}_{mn}(\bm{k})$, $R^i_{nm}(\bm{k})$ and $\Delta^i_{mn}(\bm{k})$ of the integrands in Eqs.~(\ref{eq:RLshiftcurrent}) and~(\ref{eq:RLinjectioncurrent}) are shown in Appendix~E.
Since the products are even for $\pi$-rotation, where $\Omega^{xy}_{mn}(\bm{k})$, $R^i_{nm}(\bm{k})$ and $\Delta^i_{mn}(\bm{k})$ are odd, the shift and injection current can be generated for monolayer NbSe$_2$.
However, when the products are odd for $\pi$-rotation, the current is absent.

\subsection{Selection rule by polarization}
\begin{figure}[h]
  \begin{center}
    \includegraphics[width=0.5\textwidth]{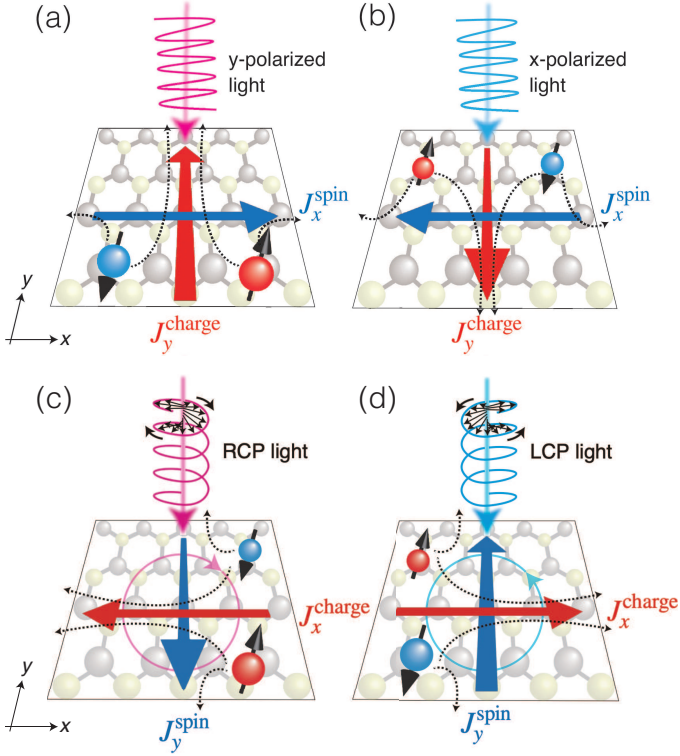}
    \caption{Schematics of generated charge and spin shift current for (a) $y$-polarized light and (b) $x$-polarized light in monolayer NbSe$_2$.
      For (c) RCP and (d) LCP light, generation of charge and spin shift current is caused in monolayer NbSe$_2$.}
    \label{fig:6}
  \end{center}
\end{figure}

The selection rules of shift- and injection-current conductivities 
by light polarization are summarized in Table~\ref{tab:table2}.
Figures~\ref{fig:6} (a) and (b) show the schematics of generated shift current in monolayer NbSe$_2$ under irradiation of LP light.
For $y$-polarized light, the charge and spin shift current can be generated in $y$- and $x$-directions.
On the other hand, for $x$-polarized light, 
the charge and spin shift current are generated in $y$- and $x$-directions, respectively.

For CP light irradiation, the charge and spin shift current can be generated in $x$- and $y$-directions as shown in Figs~\ref{fig:6} (c) and (d).
Thus, the directions of generated charge and spin shift current are switched when the polarization of incident light is altered from LP to CP. 

\section{Summary}\label{sec:level6}
In summary, we have theoretically studied 
the second-order NLO charge and spin current in the metallic TMDCs. 
As a example of metallic TMDCs, we have employed few-layered NbSe$_2$,
which possesses the Ising-type SOC.
For odd-number-layered NbSe$_2$, the inversion symmetry is broken,
however, the inversion symmetry is preserved for even-number-layered NbSe$_2$.
It is pointed out that 
the second-order NLO charge and spin current in the metallic TMDCs
strongly depends on their crystal symmetry of system and polarization of incident light. 
The NLO current is finite for odd-number-layered NbSe$_2$, but absent for even-number-layered NbSe$_2$.

Since BCL can control the symmetry of electronic system, NLO charge and
spin current can be induced along the directions of leaves of BCL. 
Thus, under irradiation of BCL, the NLO current can be generated in not
only odd-number-layered, but also even-number-layered NbSe$_2$.
For even-number-layered NbSe$_2$, the peaks appear in low energy range, which cause the infrared light absorption.
Such infrared light absorption is absent in semiconductor TMDC owing to the energy band gap around Fermi level.

In further, we have shown that shift and injection current can be generated in odd-number-layered NbSe$_2$ under irradiation of LP and CP light.
The topological shift current is robust to electron scattering of system, but the injection current strongly depends on the scattering.
The generated shift current can be switched if LP light is altered to CP light.

In Supplementary Material, the second-order NLO current can be generated even in MoS$_2$ as a reference of TMDC semiconductors~\cite{supplementary}. 
Thus, TMDCs such as NbSe$_2$ and MoS$_2$ can be used for the
source of second-order NLO charge and spin current.
Our results can serve to design spin current harvesting and opt-spintronics devices on the basis of 2D materials.

\begin{acknowledgments}
This work was supported by JSPS KAKENHI
(Nos. 22H05473, JP21H01019, JP18H01154) and JST CREST (No. JPMJCR19T1).
\end{acknowledgments}

\appendix

\section{Crystal symmetry and the second-order NLO conductivity}
\begin{figure}[h]
  \begin{center}
    \includegraphics[width=0.5\textwidth]{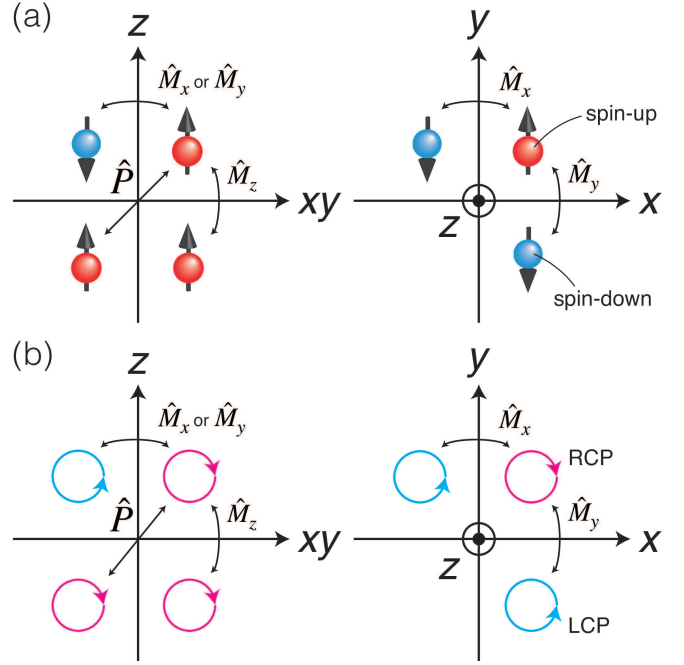}
    \caption{Symmetric operator for an axial vector.
      Schematics of mirror and inversion symmetry operations
   $\hat{M}_x$, $\hat{M}_y$, $\hat{M}_z$ and $\hat{P}$ for (a) spin-up
   (red) and spin-down (blue) states $s_z=\pm\frac{\hbar}{2}$ and (b)
   electric fields of RCP (magenta) and LCP (cyan) light.} 
    \label{fig:7}
  \end{center}
\end{figure}
The second-order NLO conductivity strongly depends on the crystal symmetry~\cite{Habara2022, Xu2021}.
Table~\ref{tab:table1} shows that the NLO charge and spin conductivities
of SHG process are finite for odd-number-layered NbSe$_2$, but absent
for even-number-layered NbSe$_2$. The finite NLO conductivities are
attributed to the broken inversion symmetry.
In this section, we discuss general NLO conductivities by considering
the crystal symmetry of system, which are consistent with the numerical
results. 

Even-number-layered NbSe$_2$ has the inversion symmetric operator $\hat{P}=\hat{M}_x\hat{M}_y\hat{M}_z$.
Here, the $3\times 3$ matrices of $\hat{M}_x$, $\hat{M}_y$ and $\hat{M}_z$ are expressed as
\begin{equation}
  \hat{M}_x=\begin{pmatrix}-1&0&0\\ 0&1&0\\ 0&0&1 \end{pmatrix}, \
	    \hat{M}_y=\begin{pmatrix}1&0&0\\ 0&-1&0\\ 0&0&1
		      \end{pmatrix}, \ \hat{M}_z=\begin{pmatrix}1&0&0\\
						 0&1&0\\ 0&0&-1
						 \end{pmatrix}.
\nonumber
\end{equation}
Because of the inversion symmetry $P$, position operator $\hat{\bm{r}}$,
momentum operator $\hat{\bm{p}}$ and spin operator
$\hat{\bm{s}}=\frac{\hbar}{2}(\hat{\sigma}_x, \hat{\sigma}_y,
\hat{\sigma}_z)$ have the following properties:
\begin{equation}
  \hat{P}\hat{\bm{r}}=-\hat{\bm{r}}, \ \hat{P}\hat{\bm{p}}=-\hat{\bm{p}}, \ \hat{P}\hat{\bm{s}}=\hat{\bm{s}}.
  \label{eq:rps}
\end{equation}
The charge current density $J^{\rm{charge}}_i$ and $j$- and $k$-elements
of incident electric fields $E_j(\omega_1)$ and $E_k(\omega_2)$ after
the operation of $\hat{P}$ become 
\begin{equation}
  \begin{split}
    J^{\rm{charge}}_i&\xrightarrow{\hat{P}} \tilde{J}^{\rm{charge}}_i=-J_i^{\rm{charge}}, \\
    E_j(\omega_1)E_k(\omega_2)&\xrightarrow{\hat{P}}
   \tilde{E}_j(\omega_1)\tilde{E}_k(\omega_2)=(-1)^2E_j(\omega_1)E_k(\omega_2), 
    \label{eq:tildecharcurele}
  \end{split}
\nonumber
\end{equation}
where $i$ is the propagation of charge current ($i=x,y,z$) and $(j, k)$ are the polarizations of incident light.
Note that $\tilde{A}$ means the arbitrary physical
quantity $A$ after the symmetry operation.
For second-order NLO response, $J^{\rm{charge}}_i$ is proportional to $E_j(\omega_1)E_k(\omega_2)$, which given as 
\begin{equation}
  \begin{split}
    J^{\rm{charge}}_i=\sigma^{\rm{charge}}_{ijk}(\omega_1+\omega_2; \omega_1, \omega_2)E_j(\omega_1)E_k(\omega_2). 
    \label{eq:curdencharijk}
  \end{split}
\end{equation}
Comparing the current and electric fields before and after $\hat{P}$ operation, the NLO conductivities become
\begin{equation}
  \sigma^{\rm{charge}}_{ijk}(\omega_1+\omega_2; \omega_1, \omega_2) =
   -\sigma^{\rm{charge}}_{ijk}(\omega_1+\omega_2; \omega_1, \omega_2).
\nonumber
\end{equation}
Thus, in even-number-layered NbSe$_2$, the NLO charge conductivity $\sigma^{\rm{charge}}_{ijk}(\omega_1+\omega_2; \omega_1, \omega_2)$ is absent for all combinations of $ijk$.

The spin current density $J^{\rm{spin}}_i$ is given as 
\begin{equation}
  \begin{split}
    J^{\rm{spin}}_i=\sigma^{\rm{spin}}_{ijk}(\omega_1+\omega_2; \omega_1, \omega_2)E_j(\omega_1)E_k(\omega_2).
    \label{eq:curdenspinijk}
  \end{split}
\end{equation}
As seen in Fig.~\ref{fig:7} (a), in even-number-layered NbSe$_2$, the spin state is invariant after $\hat{P}$.
Thus, $J^{\rm{spin}}_i$ after $\hat{P}$ becomes
\begin{equation}
  J^{\rm{spin}}_i\xrightarrow{\hat{P}}\tilde{J}^{\rm{spin}}_i=-J^{\rm{spin}}_i,
  \nonumber
\end{equation}
which causes the absence of NLO spin conductivity $\sigma^{\rm{spin}}_{ijk}(\omega_1+\omega_2; \omega_1, \omega_2)$ for even-number-layered NbSe$_2$.
These results are consistent with the numerical calculations shown in Table~\ref{tab:table1}.

In odd-number-layered NbSe$_2$, where $P$ is broken, the system has the mirror symmetry $M_xM_z$.
$J^{\rm{charge}}_i$ and $E_j(\omega_1), E_k(\omega_2)$ after the mirror symmetry operation $\hat{M}_x\hat{M}_z$ become 
\begin{equation}
  \begin{split}
    J^{\rm{charge}}_i&\xrightarrow{\hat{M}_x\hat{M}_z} \tilde{J}^{\rm{charge}}_i
    =(-1)^{\delta_{ix}+\delta_{iz}}J^{\rm{charge}}_i, \\
    E_j(\omega_1)&\xrightarrow{\hat{M}_x\hat{M}_z} \tilde{E}_j(\omega_1)=(-1)^{\delta_{jx}+\delta_{jz}}E_j(\omega_1), \\
    E_k(\omega_2)&\xrightarrow{\hat{M}_x\hat{M}_z} \tilde{E}_k(\omega_2)=(-1)^{\delta_{kx}+\delta_{kz}}E_k(\omega_2),
  \end{split}
  \nonumber
\end{equation}
where $\delta_{ix}$ is Kronecker delta: 
\begin{align}
  \delta_{ix} \ =\
  \begin{cases}
    1 & (i=x) \\
    0 & (i\not=x)
  \end{cases}.
\end{align}
Thus, $\hat{M}_x\hat{M}_z$ makes the following $ijk$-elements of $\sigma^{\rm{charge}}_{ijk}(\omega_1+\omega_2; \omega_1, \omega_2)$ in Eq.~(\ref{eq:curdencharijk}) finite, i.e., $ijk=xxy, xyx, xyz, xzy, yxx, yxz, yyy, yzx, yzz, zxy, zyx, zyz, zzy$.
The other elements are absent for odd-number-layered NbSe$_2$.
In addition, the system has $3$-fold rotation symmetry operation $\hat{C}_3$, which is given as 
\begin{equation}
  \hat{C}_3=\begin{pmatrix}
    -\frac{1}{2}&\frac{\sqrt{3}}{2}&0 \\
    -\frac{\sqrt{3}}{2}&-\frac{1}{2}&0 \\
    0&0&1
  \end{pmatrix}.
  \nonumber
\end{equation}
Owing to $\hat{C}_3$, the finite NLO charge conductivities have a relation for odd-number-layered NbSe$_2$, i.e., 
\begin{equation}
  \begin{split}
    &\sigma^{\rm{charge}}_{yyy}(\omega_1+\omega_2; \omega_1, \omega_2)=-\sigma^{\rm{charge}}_{xxy}(\omega_1+\omega_2; \omega_1, \omega_2)\\
    &=-\sigma^{\rm{charge}}_{xyx}(\omega_1+\omega_2; \omega_1, \omega_2)=-\sigma^{\rm{charge}}_{yxx}(\omega_1+\omega_2; \omega_1, \omega_2).
  \end{split}
\end{equation}

In Fig.~\ref{fig:7} (a), the spin state becomes opposite after $\hat{M}_x\hat{M}_z$, which causes
\begin{equation}
  J^{\rm{spin}}_i\xrightarrow{\hat{M}_x\hat{M}_z}\tilde{J}^{\rm{spin}}_i=-(-1)^{\delta_{ix}+\delta_{iz}}J^{\rm{spin}}_i.
  \nonumber
\end{equation}
Thus, the following $ijk$-elements of $\sigma^{\rm{spin}}_{ijk}(\omega_1+\omega_2; \omega_1, \omega_2)$ in Eq.~(\ref{eq:curdenspinijk}) are finite for odd-number-layered NbSe$_2$: $ijk=xxx, xxz, xyy, xzx, xzz, yxy, yyx, yyz, yzy, zxx, zxz, zyy, zzx, zzz$.
The other elements are absent.
Also, for odd-number-layered NbSe$_2$, the following relation of $\sigma^{\rm{spin}}_{ijk}(\omega_1+\omega_2; \omega_1, \omega_2)$ is obtained after $\hat{C}_3$, i.e., 
\begin{equation}
  \begin{split}
    \sigma^{\rm{spin}}_{xxx}&(\omega_1+\omega_2; \omega_1, \omega_2)=-\sigma^{\rm{spin}}_{xyy}(\omega_1+\omega_2; \omega_1, \omega_2)\\
    &=-\sigma^{\rm{spin}}_{yxy}(\omega_1+\omega_2; \omega_1, \omega_2)=-\sigma^{\rm{spin}}_{yyx}(\omega_1+\omega_2; \omega_1, \omega_2).
  \end{split}
\end{equation}

These results indicate that spin current can be generated in odd-number-layered NbSe$_2$ when charge current is absent.
When $\omega_1=\omega_2\equiv\omega\not=0$, the analytic results are consistent with the numerical results of SHG process shown in Table~\ref{tab:table1}.

Next, by considering the crystal symmetry, we shall discuss the generation of NLO charge and spin current in NbSe$_2$
under CP light irradiation.
For CP light, $J^{\rm{charge}}_i$ is given as 
\begin{equation}
  \begin{split}
    J^{\rm{charge}}_i=\sigma^{\rm{charge}}_{iRL}(\omega_1+\omega_2; \omega_1, \omega_2)E_R(\omega_1)E_L(\omega_2).
  \end{split}
\end{equation}
For even-number-layered NbSe$_2$, the product of electric fields of CP light $E_R(\omega_1)E_L(\omega_2)$ after $\hat{P}$ becomes 
\begin{equation}
  \begin{split}
    E_R(\omega_1)E_L(\omega_2)&\xrightarrow{\hat{P}} \tilde{E}_R(\omega_1)\tilde{E}_L(\omega_2)=E_R(\omega_1)E_L(\omega_2), 
  \end{split}
  \nonumber
\end{equation}
where as shown in Fig.~\ref{fig:7} (b), $E_R(\omega_1), E_L(\omega_2)$ are invariant for $\hat{P}$.
Comparing the current and the electric fields of incident light before and after $\hat{P}$, $\sigma^{\rm{charge}}_{iRL}(\omega_1+\omega_2; \omega_1, \omega_2)$ becomes zero.
Since $J^{\rm{spin}}_i$ is invariant after $\hat{P}$, $\sigma^{\rm{spin}}_{iRL}(\omega_1+\omega_2; \omega_1, \omega_2)$ is absent as same as NLO charge conductivity.

In odd-number-layered NbSe$_2$, $E_R(\omega_1)E_L(\omega_2)$ after $\hat{M}_x\hat{M}_z$ becomes 
\begin{equation}
  \begin{split}
    E_R(\omega_1)E_L(\omega_2)&\xrightarrow{\hat{M}_x\hat{M}_z} \tilde{E}_R(\omega_1)\tilde{E}_L(\omega_2)=E_L(\omega_1)E_R(\omega_2).
  \end{split}
  \label{eq:charmxmzJE}
  \nonumber
\end{equation}
Thus, $\hat{M}_x\hat{M}_z$ makes $\sigma^{\rm{charge}}_{xRL}(\omega_1+\omega_2; \omega_1, \omega_2)$ and $\sigma^{\rm{charge}}_{zRL}(\omega_1+\omega_2; \omega_1, \omega_2)$ finite, but $\sigma^{\rm{charge}}_{yRL}(\omega_1+\omega_2; \omega_1, \omega_2)$ is absent.
Since the spin state becomes opposite owing to $\hat{M}_x\hat{M}_z$, the NLO spin conductivity $\sigma^{\rm{spin}}_{yRL}(\omega_1+\omega_2; \omega_1, \omega_2)$ is finite, but the other elements are absent.
These results are consistent with the numerical calculation summarized in Table~\ref{tab:table2}.

In further, under irradiation of BCL ($A_{\rm{BCL}}(t)=A_Le^{in_1\omega t}+A_Re^{in_2\omega t-i\theta}+{\rm c.c.}$), the mirror symmetry $M_x$ is broken.
Thus, $\hat{M}_y\hat{M}_z$ remains in the electronic system. 
$J^{\rm{charge}}_i$ and $E_{\rm{BCL}}(t)$ after $\hat{M}_y\hat{M}_z$ become 
\begin{equation}
  \begin{split}
    J^{\rm{charge}}_i&\xrightarrow{\hat{M}_y\hat{M}_z} \tilde{J}^{\rm{charge}}_i=(-1)^{\delta_{iy}+\delta_{iz}}J^{\rm{charge}}_i, \\
    E_{\rm{BCL}}(t)&\xrightarrow{\hat{M}_y\hat{M}_z} \tilde{E}_{\rm{BCL}}(t)=E_{\rm{BCL}}(t).
  \end{split}
  \nonumber
\end{equation}
Comparing the current and the electric field of incident light before and after $\hat{M}_y\hat{M}_z$, the NLO charge conductivity $\sigma^{\rm{charge}}_{x-\rm{BCL}-\rm{BCL}}(n_1\omega, n_2\omega)$ becomes finite.
However, the other elements are absent for even-number-layered NbSe$_2$.
As seen in Fig.~\ref{fig:7} (a), for $\hat{M}_y\hat{M}_z$, the spin state becomes opposite. 
Thus, $\sigma^{\rm{spin}}_{y-\rm{BCL}-\rm{BCL}}(n_1\omega, n_2\omega)$ and $\sigma^{\rm{spin}}_{z-\rm{BCL}-\rm{BCL}}(n_1\omega, n_2\omega)$ are finite, but $\sigma^{\rm{spin}}_{x-\rm{BCL}-\rm{BCL}}(n_1\omega, n_2\omega)$ is absent.
These analytic results show that thanks to incident BCL, the NLO charge and spin current can be generated even in even-number-layered NbSe$_2$.

Odd-number-layered NbSe$_2$ has $M_xM_z$.
Since $M_x$ is broken under BCL irradiation, the electronic system has $\hat{M}_z$.
For $\hat{M}_z$, $J^{\rm{charge}}_i$ and $E_{\rm{BCL}}(t)$ become 
\begin{equation}
  \begin{split}
    J^{\rm{charge}}_i&\xrightarrow{\hat{M}_z} \tilde{J}^{\rm{charge}}_i=(-1)^{\delta_{iz}}J^{\rm{charge}}_i, \\
    E_{\rm{BCL}}(t)&\xrightarrow{\hat{M}_z} \tilde{E}_{\rm{BCL}}(t)=E_{\rm{BCL}}(t).
  \end{split}
  \nonumber
\end{equation}
Owing to the comparison between the generated current and incident light before and after $\hat{M}_z$, $\sigma^{\rm{charge}}_{x-\rm{BCL}-\rm{BCL}}(n_1\omega, n_2\omega)$ and $\sigma^{\rm{charge}}_{y-\rm{BCL}-\rm{BCL}}(n_1\omega, n_2\omega)$ become finite for odd-number-layered NbSe$_2$, but $\sigma^{\rm{charge}}_{z-\rm{BCL}-\rm{BCL}}(n_1\omega, n_2\omega)$ is absent.
Since the spin state is invariant after $\hat{M}_z$, $\sigma^{\rm{spin}}_{x-\rm{BCL}-\rm{BCL}}(n_1\omega, n_2\omega)$ and $\sigma^{\rm{spin}}_{y-\rm{BCL}-\rm{BCL}}(n_1\omega, n_2\omega)$ are finite, but $\sigma^{\rm{spin}}_{z-\rm{BCL}-\rm{BCL}}(n_1\omega, n_2\omega)$ is absent.

\section{NLO charge and spin conductivities for irradiation of BCL with multi-leaves}
\begin{figure*}[t]
  \begin{center}
    \includegraphics[width=0.90\textwidth]{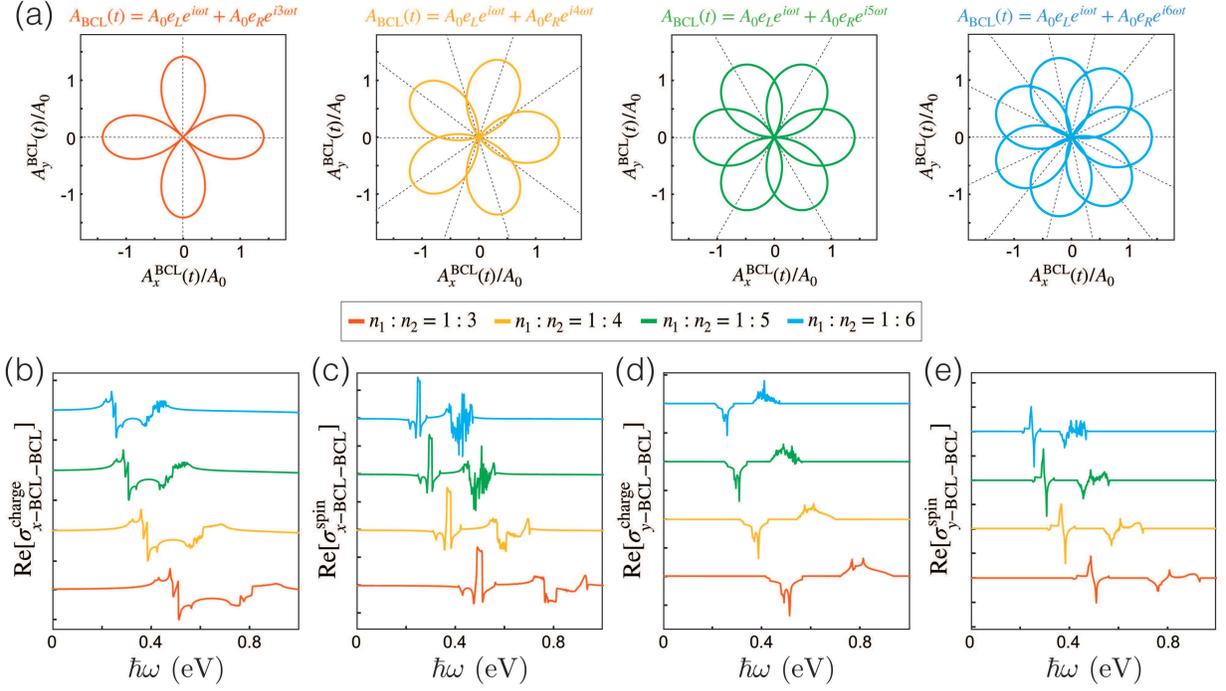}
    \caption{(a) Trajectories of BCL with $4$-, $5$-, $6$- and $7$-leaf.
      Real parts of NLO charge and spin conductivities (b) $\sigma^{\rm{charge}}_{x-\rm{BCL}-\rm{BCL}}$, (c) $\sigma^{\rm{spin}}_{x-\rm{BCL}-\rm{BCL}}$, (d) $\sigma^{\rm{charge}}_{y-\rm{BCL}-\rm{BCL}}$ and (e) $\sigma^{\rm{spin}}_{y-\rm{BCL}-\rm{BCL}}$ for monolayer NbSe$_2$ with SOC under irradiation of BCL.
      Orange, yellow, green and blue lines indicate NLO conductivities for BCL with $4$-, $5$-, $6$- and $7$-leaf.
      The units of NLO charge and spin conductivities are $e^3/\hbar$ and $e^2$, respectively.}
    \label{fig:8}
  \end{center}
\end{figure*}
In this section, we consider BCL of $4$-, $5$-, $6$- and $7$-leaf with $\theta=0$ in Eq.~(\ref{eq:bcl}).
Figure~\ref{fig:8} (a) shows the trajectories of these incident BCL.
Figures~\ref{fig:8} (b), (c), (d) and (e) show the NLO charge and spin conductivities for monolayer NbSe$_2$ with SOC under irradiation of BCL.
Since the BCL can control the symmetry of electronic system, these NLO charge and spin conductivities are finite for monolayer NbSe$_2$, which are similar to the results of BCL with $3$-leaf ($n_1:n_2=1:2$).
Thus, unlike the irradiation of LP light, the NLO charge and spin current is generated in both $x$- and $y$-directions along the induced polarizations by irradiating BCL.

We should note that the peaks of the NLO charge and spin conductivities shift lower with increase of leaves of BCL.
The shift originates from the denominator
$E_{mn}-(n_1+n_2)\hbar\omega-i\eta$ in Eq.~(\ref{eq:2ndkubo2}).
With increase of $(n_1+n_2)$, the denominator $E_{mn}-(n_1+n_2)\hbar\omega-i\eta$ becomes smaller, thus the shift occurs.

\section{Finite conductivity for CP light with different optical angular frequency}

\begin{figure*}[t]
  \begin{center}
    \includegraphics[width=1.0\textwidth]{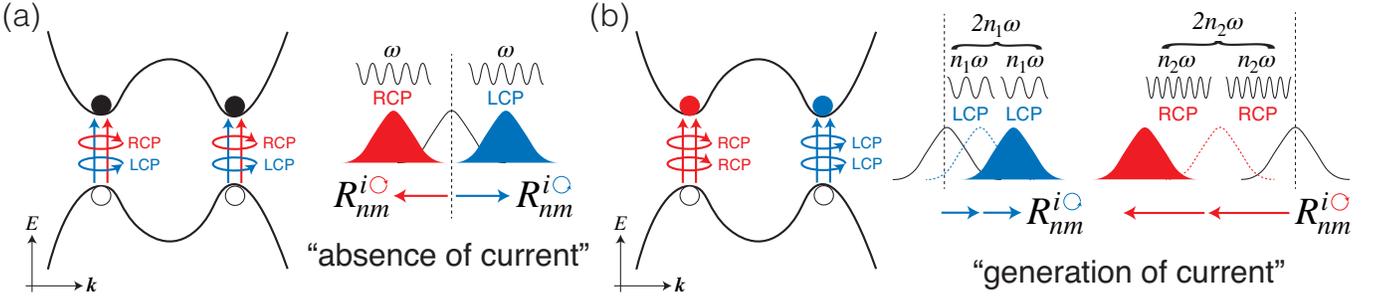}
    \caption{Schematics of photoexcitation for CP light irradiation and
   corresponding shift vectors.
      (a) For irradiation of RCP (red) and LCP (blue) light with same
   optical angular frequency $\omega$,
RCP and LCP make same amount of shift but opposite direction,
   i.e., absence of current. 
      (b) For irradiation of RCP and LCP light with different optical
   angular frequencies $n_1\omega\not=n_2\omega$,
the amount of shifts by RCP and LCP is different and no cancellation, i.e., generation of current.}
    \label{fig:9}
  \end{center}
\end{figure*}

The shift current under irradiation of CP light contains
the product of Berry curvature $\Omega^{xy}_{mn}(\bm{k})$ and shift
vector $R^i_{nm}(\bm{k})$ as shown in Eq.~(\ref{eq:RLshiftcurrent}).
The Berry curvatures between RCP and LCP light are
antisymmetric, thus the shift current for irradiation of RCP and LCP
light is generated in opposite directions.

When the optical angular frequencies of RCP and LCP light are same, as
shown in Fig.~\ref{fig:9} (a), 
the cancellation of generated shift current occurs for RCP and LCP
light, i.e., absence of current~\cite{Ikeda2022, Watanabe2021, Ma2023}. 
Figure~\ref{fig:9} (b) shows the schematics of shift current 
under the irradiation of LCP light with $2n_1\omega$ and RCP light with
$2n_2\omega$. 
If LCP and RCP light has the different optical angular
frequencies, i.e., $n_1\not=n_2$, the amounts of shifts for photoexcited
electrons under irradiation of LCP and RCP light are not same, 
i.e., no cancellation of shift current. 

In Eq.~(\ref{eq:curdensbcl}), the NLO conductivity for BCL irradiation
is rewritten as the summation of conductivities for different four
combinations of CP light, i.e., `LL', `LR', `RL', `RR'. 
Owing to the irradiation of CP light with different optical angular frequencies, the first and fourth terms are finite, but other terms are cancelled.
Thus, for irradiation of BCL, the second-order NLO current is generated even in even-number-layered NbSe$_2$.

\section{NLO charge and spin current in trilayer NbSe$_2$ under irradiation of BCL}
\begin{figure*}[t]
  \begin{center}
    \includegraphics[width=0.90\textwidth]{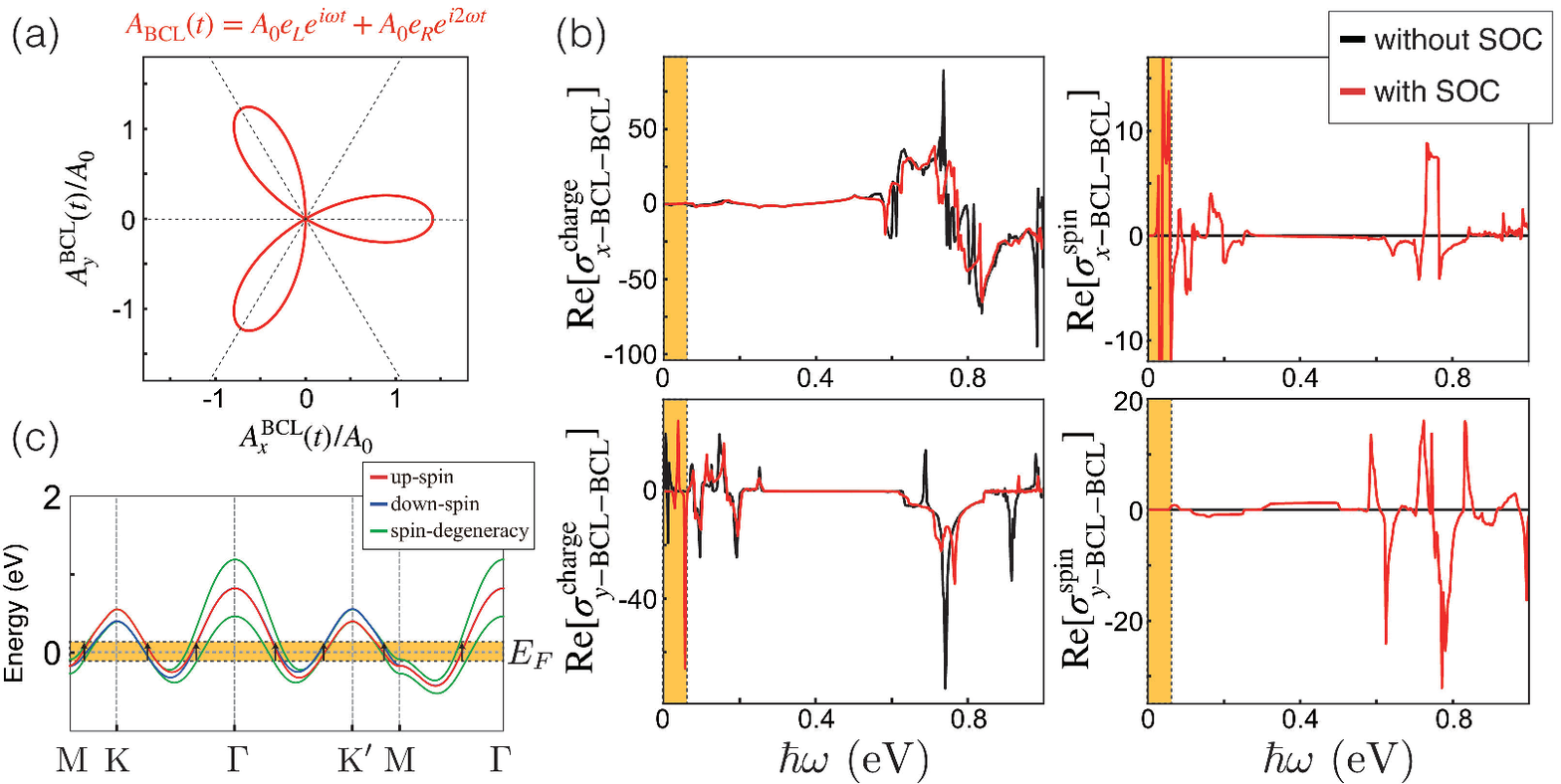}
    \caption{(a) Trajectory of BCL with $3$-leaf ($n_1:n_2=1:2$).
      (b) Real parts of NLO charge and spin conductivities $\sigma^{\rm{charge}}_{x-\rm{BCL}-\rm{BCL}}$, $\sigma^{\rm{spin}}_{x-\rm{BCL}-\rm{BCL}}$, $\sigma^{\rm{charge}}_{y-\rm{BCL}-\rm{BCL}}$ and $\sigma^{\rm{spin}}_{y-\rm{BCL}-\rm{BCL}}$ for trilayer NbSe$_2$.
      Black and red lines indicate NLO conductivities without and with SOC.
      The units of NLO charge and spin conductivities are $e^3/\hbar$ and $e^2$, respectively.
      (c) Energy band structure of trilayer NbSe$_2$ and the interband optical transition between nearest valence and conduction bands around Fermi energy $E_F=0$ eV (shown by a yellow square).
    }
    \label{fig:10}
  \end{center}
\end{figure*}
As shown in Fig.~\ref{fig:10} (a), we consider $n_1:n_2=1:2$ and $\theta=0$ for BCL with $3$-leaf.
Figure~\ref{fig:10} (b) shows the NLO charge and spin conductivities for trilayer NbSe$_2$ under irradiation of the BCL.
Since the BCL breaks the mirror symmetry with respect to $y-z$ plane, these NLO charge and spin conductivities are finite for trilayer NbSe$_2$.
Here, it should be noted that the peaks of the NLO conductivities appear around $\hbar\omega=0.01$ eV shown by yellow squares. 
The appearance of peaks occurs owing to the photoexcitation between the nearest valence and conduction bands of trilayer NbSe$_2$ as shown in Fig.~\ref{fig:10} (c), which causes infrared light absorption.

In further, for trilayer NbSe$_2$, the NLO conductivities under irradiation of $3$-leaf BCL consist of three optical transition processes: (i) intralayer optical transition of monolayer NbSe$_2$, (ii) intra- and interlayer optical transitions of bilayer NbSe$_2$ and (iii) interlayer optical transition between monolayer and bilayer NbSe$_2$.
For irradiation of LP light, the process (ii) is absent~\cite{Habara2022}, however BCL with $3$-leaf makes all three processes finite.

\section{Contour plots of $\alpha^{jk}_{mn}(\bm{k})$, $R^i_{nm}(\bm{k})$, $\Delta^{i}_{mn}(\bm{k})$, $\Omega^{jk}_{mn}(\bm{k})$}
\begin{figure*}[t]
  \begin{center}
    \includegraphics[width=0.85\textwidth]{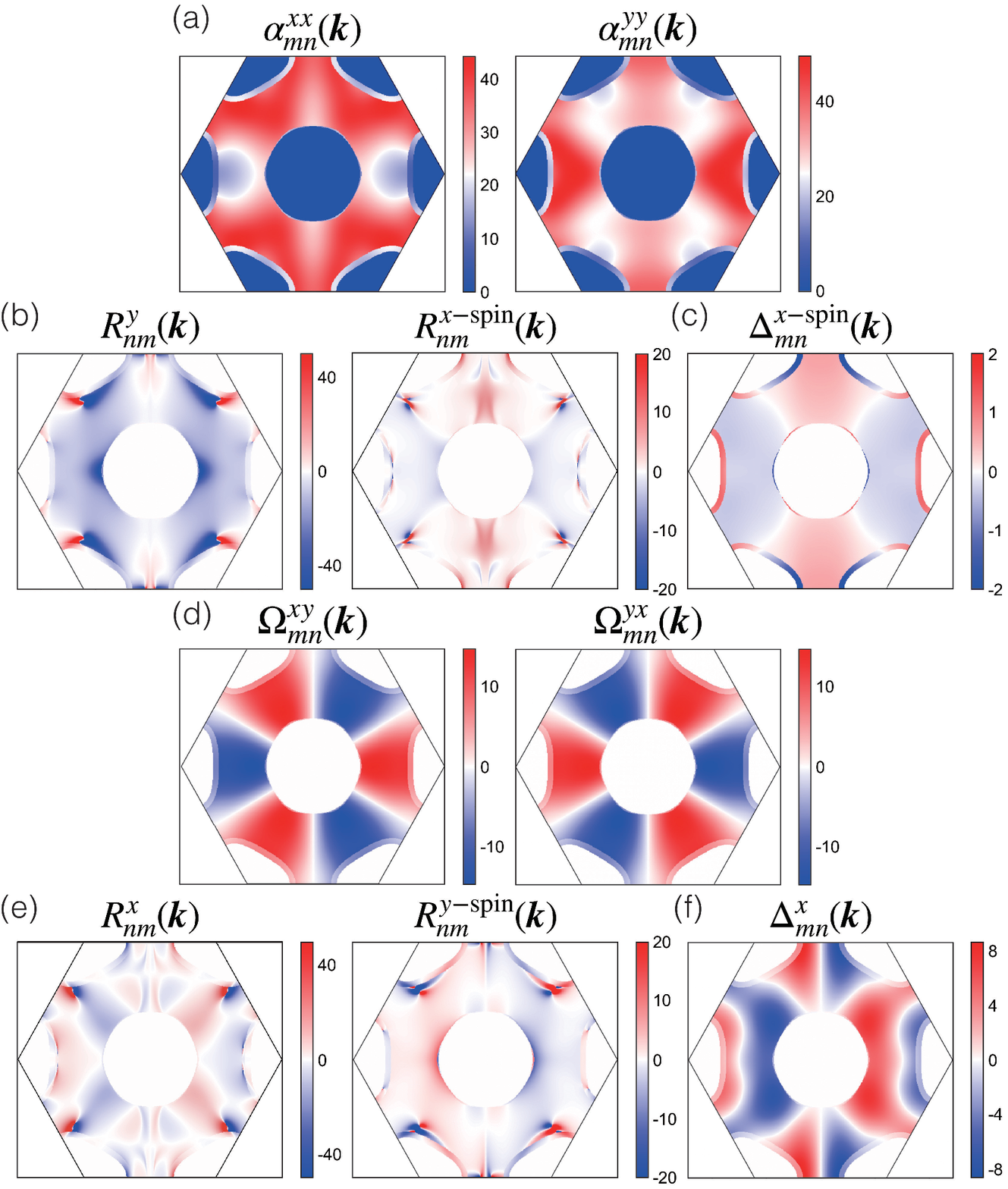}
    \caption{Contour plots of shift- and injection-current conductivity components: (a) transition intensities $\alpha^{xx}_{mn}(\bm{k})$ and $\alpha^{yy}_{mn}(\bm{k})$, (b, e) shift vectors $R^y_{nm}(\bm{k})$, $R^{x-\rm{spin}}_{nm}(\bm{k})$, $R^x_{nm}(\bm{k})$ and $R^{y-\rm{spin}}_{nm}(\bm{k})$, (c, f) velocity difference between electron and hole $\Delta^{x-\rm{spin}}_{mn}(\bm{k})$ and $\Delta^x_{mn}(\bm{k})$, and (d) Berry curvatures $\Omega^{xy}_{mn}(\bm{k})$ and $\Omega^{yx}_{mn}(\bm{k})$, respectively.}
    \label{fig:11}
  \end{center}
\end{figure*}
We discuss the origin of finite shift and injection current by
considering the contour plots of transition intensity
$\alpha^{jk}_{mn}(\bm{k})$, shift vector $R^i_{nm}(\bm{k})$, velocity difference
between electron and hole $\Delta^i_{mn}(\bm{k})$, and Berry
curvature $\Omega^{jk}_{mn}(\bm{k})$.
All these factors appear as the integrand of the conductivities. 
For simplicity, we shall focus on the results of monolayer NbSe$_2$,
because other cases such bi- and trilayer NbSe$_2$ are similar.

In Eq.~(\ref{eq:shiftcur}), the charge (spin) shift-current conductivity for LP light
contains
the product $\alpha^{jk}_{mn}(\bm{k})R^i_{nm}(\bm{k})$ ($\alpha^{jk}_{mn}(\bm{k})R^{i-\rm{spin}}_{nm}(\bm{k})$).
If the product $\alpha^{jk}_{mn}(\bm{k})R^i_{nm}(\bm{k})$ 
is even with respect to $\pi$-rotation, the shift-current
conductivities become finite. 
According to the contour plots of transition intensity shown in Fig.~\ref{fig:11} (a),
$\alpha^{jk}_{mn}(\bm{k})$ with $jk=xx, yy$ is even for $\pi$-rotation.
Similarly, 
as shown in Fig.~\ref{fig:11} (b),
the shift vectors 
$R^y_{nm}(\bm{k})$ and $R^{x-\rm{spin}}_{nm}(\bm{k})$ are also even for $\pi$-rotation.
Therefore, the charge and spin shift current are generated in $y$- and
$x$-directions, respectively. 
The remaining elements $R^x_{nm}(\bm{k})$ and
$R^{y-\rm{spin}}_{nm}(\bm{k})$ are odd for $\pi$-rotation, i.e.,
the absence of shift current.


In Eq.~(\ref{eq:injectcur}), the charge (spin) injection-current
 conductivity for LP light 
contains $\alpha^{jk}_{mn}(\bm{k})\Delta^i_{mn}(\bm{k})$
 ($\alpha^{jk}_{mn}(\bm{k})\Delta^{i-\rm{spin}}_{mn}(\bm{k})$). 
Since $\alpha^{jk}_{mn}(\bm{k})$ with $jk=xx, yy$ is even with respect
 to $\pi$-rotation, 
velocity difference
$\Delta^{i}_{mn}(\bm{k})$ should be even in order to generate the injection current.
In actual, as shown in Fig.~\ref{fig:11} (c), 
$\Delta^{x-\rm{spin}}_{mn}(\bm{k})$ has even parity.
Thus, the spin injection current is generated in $x$-direction.
The remaining elements of $\Delta^{i}_{mn}(\bm{k})$ are odd, i.e., the absence of injection current.

Similarly, in Eqs.~(\ref{eq:RLshiftcurrent})
and~(\ref{eq:RLinjectioncurrent}), the charge (spin)
shift/injection-current conductivities for RCP light contain the
products $\Omega^{xy}_{mn}(\bm{k})R^i_{nm}(\bm{k})$
($\Omega^{xy}_{mn}(\bm{k})R^{i-\rm{spin}}_{nm}(\bm{k})$) and
$\Omega^{xy}_{mn}(\bm{k})\Delta^i_{mn}(\bm{k})$
($\Omega^{xy}_{mn}(\bm{k})\Delta^{i-\rm{spin}}_{mn}(\bm{k})$),
respectively. 
When the products are even with respect to $\pi$-rotation, the shift and
injection current is generated for RCP light. 
Figures~\ref{fig:11} (d), (e) and (f) show the contour plots of Berry
curvature, shift vector and velocity difference for monolayer NbSe$_2$ under RCP light
irradiation. 
From Fig.~\ref{fig:11} (d), $\Omega^{xy}_{mn}(\bm{k})$ and $\Omega^{yx}_{mn}(\bm{k})$ are odd for
$\pi$-rotation.
Since $R^x_{nm}(\bm{k})$ and $R^{y-\rm{spin}}_{nm}(\bm{k})$ are odd
parities, the charge and spin shift current is generated in $x$- and
$y$-directions, respectively. Similarly, since 
$\Delta^{x}_{mn}(\bm{k})$ is odd parity, the charge injection current is
generated in $x$-direction.

\nocite{*}
\bibliography{reference}
\end{document}